\documentclass[aps,prl, nofootinbib, twocolumn,superscriptaddress]{revtex4-2}
\usepackage{subfigure}
\usepackage{graphicx}
\usepackage{dcolumn}
\usepackage{bm}
\usepackage{xcolor}
\usepackage{float}
\definecolor{aa}{RGB}{0,0,139}

\newcommand{\edos}{$e^+e^-\to D_s^+D_{s1}(2536)^-$}
\newcommand{\dos}{$D_{s1}(2536)^-\to \bar{D}^{*0}K^-$}
\newcommand{\edts}{$e^+e^-\to D_s^+D^*_{s2}(2573)^-$}

\newcommand{\kkpi}{$D_s^+\to K^-K^+\pi^+$}
\newcommand{\ksk}{$D_s^+\to K_S^0(\to \pi^+\pi^-)K^+$}
\newcommand{\FS}[2]{\displaystyle\frac{#1}{#2}}
\newcommand{\dskkp}{D_s^+ \to K^-K^+\pi^+}

\newcommand{\dsksk}{D_s^+\to K_S^0K^+}
\newcommand{\dod}{D_{s1}(2536)^- \to \bar{D}^{*0}K^-}
\newcommand{\dtd}{D_{s2}^*(2573)^- \to \bar{D}^0K^-}

\usepackage[colorlinks,urlcolor=aa,linkcolor=blue,anchorcolor=aa,citecolor=aa]{hyperref}
\usepackage[mathlines]{lineno}
\usepackage{multirow}
\usepackage{enumerate}
\usepackage{amsmath}
\usepackage{relsize}
\def\babar{\mbox{\slshape B\kern-0.1em{\smaller A}\kern-0.1em
    B\kern-0.1em{\smaller A\kern-0.2em R}}}

\begin{document}

\preprint{APS/123-QED}

\title{\boldmath Study of the decay and production properties of $D_{s1}(2536)$ and $D_{s2}^*(2573)$}

\author{
M.~Ablikim$^{1}$, M.~N.~Achasov$^{4,c}$, P.~Adlarson$^{76}$, O.~Afedulidis$^{3}$, X.~C.~Ai$^{81}$, R.~Aliberti$^{35}$, A.~Amoroso$^{75A,75C}$, Q.~An$^{72,58,a}$, Y.~Bai$^{57}$, O.~Bakina$^{36}$
% ...
\\
\vspace{0.2cm}
(BESIII Collaboration)\\
\vspace{0.2cm} {\it
$^{1}$ Institute of High Energy Physics, Beijing 100049, People's Republic of China\\
% ...
$^{3}$ Bochum  Ruhr-University, D-44780 Bochum, Germany\\
$^{4}$ Budker Institute of Nuclear Physics SB RAS (BINP), Novosibirsk 630090, Russia\\
% ...
$^{35}$ Johannes Gutenberg University of Mainz, Johann-Joachim-Becher-Weg 45, D-55099 Mainz, Germany\\
$^{36}$ Joint Institute for Nuclear Research, 141980 Dubna, Moscow region, Russia\\
% ...
$^{57}$ Southeast University, Nanjing 211100, People's Republic of China\\
$^{58}$ State Key Laboratory of Particle Detection and Electronics, Beijing 100049, Hefei 230026, People's Republic of China\\
% ...
$^{72}$ University of Science and Technology of China, Hefei 230026, People's Republic of China\\
% ...
$^{75}$ University of Turin and INFN, (A)University of Turin, I-10125, Turin, Italy; (B)University of Eastern Piedmont, I-15121, Alessandria, Italy; (C)INFN, I-10125, Turin, Italy\\
$^{76}$ Uppsala University, Box 516, SE-75120 Uppsala, Sweden\\
% ...
$^{81}$ Zhengzhou University, Zhengzhou 450001, People's Republic of China\\
\vspace{0.2cm}
$^{a}$ Deceased\\
% ...
$^{c}$ Also at the Novosibirsk State University, Novosibirsk, 630090, Russia\\
% ...
}}

\begin{abstract}

The \edos\ and \edts\ processes are studied using data samples
collected with the BESIII detector at center-of-mass energies from
4.530 to 4.946~GeV.  The absolute branching fractions of $\dod$ and
$\dtd$ are measured for the first time to be $(35.9\pm 4.8\pm 3.5)\%$
and $(37.4\pm 3.1\pm 4.6)\%$, respectively.  The measurements are in
tension with predictions based on the assumption that the
$D_{s1}(2536)$ and $D_{s2}^*(2573)$ are dominated by a bare $c\bar{s}$
component.  The \edos\ and \edts\ cross sections are measured, and a
resonant structure at around 4.6~GeV with a width of 50~MeV is
observed for the first time with a statistical significance of
$15\sigma$ in the \edts\ process. It could be the $Y(4626)$ found by
the Belle collaboration in the $D_s^+D_{s1}(2536)^{-}$ final state,
since they have similar masses and widths.  There is also evidence for
a structure at around 4.75~GeV in both processes.

\end{abstract}



\maketitle

The $D_s$ mesons are bound states of $c\bar{s}$ quarks.  Four $P$-wave
$c\bar{s}$ states with $J^{P}=0^+$ ($D_{s0}^*$), $1^+$ ($D_{s1}$),
$1^+$ ($D_{s1}^{\prime}$), and $2^+$ ($D_{s2}^*$) are predicted in the
conventional quark model \cite{frac_Dsj_ex}, and the four
experimentally observed states $D^*_{s0}(2317)$, $D_{s1}(2460)$,
$D_{s1}(2536)$, and $D_{s2}^*(2573)$ are assigned to them,
respectively.  Recently, authors of Ref.~\cite{inner_Dsj} developed a
coupled-channel framework which considers the quark-pair-creation
mechanism and $D^{(*)}K$ interactions to investigate the inner
structures of these states.  The framework explains the lower measured
masses of $D^*_{s0}(2317)$~\cite{Ds0_2317} and
$D_{s1}(2460)$~\cite{Ds1_2460} compared with those predicted by the
conventional quark model and infers that $(98.2^{+0.1}_{-0.2})\%$ and
$(95.9^{+1.0}_{-1.5})\%$ of the contents of the $D_{s1}(2536)$ and
$D_{s2}^*(2573)$, respectively, are bare $c\bar{s}$
cores~\cite{inner_Dsj}.  At the heavy quark limit and regarding
$D_{s1}(2536)$ and $D_{s2}^*(2573)$ as being dominated by a bare
$c\bar{s}$ core, authors of Ref.~\cite{frac_Dsj_ex} predict the
absolute branching fractions of $D_{s1}(2536)\to D^*K$ and
$D_{s2}^*(2573)\to DK$ to be 100\% and 93.4\%,
respectively. Experimental measurements of $D_{s1}(2536)\to D^*K$ and
$D_{s2}^*(2573)\to DK$ play an important role in understanding the
inner structure of these $P$-wave charmed-strange mesons.

Effective Field Theory~\cite{QCD1, QCD2, QCD3} and Quantum
Chromodynamics-inspired potential models~\cite{QCDPM1, QCDPM2, QCDPM3,
  QCDPM4} predict six vector charmonium states with masses between
4.0 and 4.8~GeV/$c^2$: $\psi(3^3S_1)$, $\psi(2^3D_1)$, $\psi(4^3S_1)$,
$\psi(3^3D_1)$, $\psi(5^3S_1)$, and $\psi(4^3D_1)$. The first three
states are usually assigned as $\psi(4040)$, $\psi(4160)$, and
$\psi(4415)$, respectively.  The unclassified $\psi(3^3D_1)$,
$\psi(5^3S_1)$, and $\psi(4^3D_1)$ states are expected to have masses
above 4.45~GeV/$c^2$. However, the $Y(4500)$~\cite{Y4500KKJpsiBESIII},
$Y(4660)$~\cite{ISRPipiPsipBelle}, $Y(4710)$~\cite{Y4710KKJpsiBESIII},
and $Y(4790)$~\cite{Y4790KKJpsiBESIII} are observed in this mass
region, which makes the assignment of these states very uncertain.
The $Y(4660)$ is observed through initial state radiation (ISR) in
$e^+e^-\to \pi^+\pi^-\psi(2S)$~\cite{ISRPipiPsipBelle}, and the
$\pi^+\pi^-$ invariant mass tends to accumulate at the nominal mass of
$f_{0}(980)$, which has an $s\bar{s}$
component~\cite{ISRPipiPsipBelle}; the $Y(4500)$ and $Y(4710)$ are
observed in $e^+e^-\to K^+K^-J/\psi$~\cite{Y4500KKJpsiBESIII,
  Y4710KKJpsiBESIII} and the $Y(4790)$ in $e^+e^-\to
D_s^{*+}D_s^{*-}$~\cite{Y4790KKJpsiBESIII}, where $K$ and $D_s^*$ have
 $s$ components also.  These measurements indicate that these four
states have both $s\bar{s}$ and $c\bar{c}$ components and may decay
into a charmed-strange meson pair. Therefore, the search for possible
$Y$ states in $c\bar{s}$ and $\bar{c}s$ meson pairs provides an
opportunity to investigate these unclassified $Y$ states.  Evidence
for $Y(4626)$ in $e^+e^-\to D_s^+D_{s1}(2536)^-$ (charge conjugated
processes and particles are always implied in the
following)~\cite{DsDs2536Belle} and evidence for a $Y(4620)$ state in
$e^+e^-\to D_s^+D^*_{s2}(2573)^-$~\cite{DsDs2573Belle} are reported by
the Belle collaboration in ISR processes with large
uncertainties. Improved measurements at BESIII and other experiments
are needed to draw more solid conclusions on these states.

In this Letter, the \edos\ and $D_s^+D_{s2}^*(2573)^{-}$ processes are
investigated with $D_{s1}(2536)^-$ and $D^*_{s2}(2573)^-$ decaying
both inclusively (inclusive analysis) and to $\bar{D}^{*0}K^-$ and
$\bar{D}^0K^-$ (exclusive analysis). The absolute
branching fractions of $\dod$ and $\dtd$ are measured by comparing the
cross sections of inclusive and exclusive processes, and possible $Y$
states are searched for in the exclusive cross sections.

The BESIII detector is described in detail in
Refs.~\cite{BESIIIdetector, detvis}.  The experimental data samples
used in this Letter are taken at center-of-mass energies ($\sqrt{s}$)
ranging from $4.530$ to $4.946$~GeV with 15 energy
points~\cite{XYZEcm1,XYZEcm2} corresponding in total to an integrated
luminosity of $6.60~\textrm{fb}^{-1}$~\cite{Lum1,XYZEcm2}; the details
of the data samples are shown in the supplemental material.  Since the
cross sections of some background processes are not measured for data
samples with $\sqrt{s}<4.6$~GeV and $\sqrt{s}>4.7$~GeV, only data
samples with $4.6\leq\sqrt{s}<4.74$~GeV (excluding
$\sqrt{s}=4.610$~GeV due to low statistics) are used for the absolute
branching fraction measurements. Cross sections of the exclusive
processes at all energy points are measured.  Simulated samples, which
are used to estimate the background and to determine the detection
efficiencies and ISR correction factors, are produced with {\sc
  geant4}-based~\cite{geant4} Monte Carlo (MC) software, which
includes the geometric description of the BESIII detector and its
response.

The \edos\ process is simulated with the {\sc AngSam}
model~\cite{simulation1,simulation2}, using an angular distribution
described by $1+\alpha {\rm cos}^2\theta$, where $\theta$ is the polar
angle of $D_s^+$ in the $e^+e^-$ rest frame, and $\alpha=-0.65\pm
0.22$ is measured in this work.  The \dos\ decay is simulated with the
VVS\_PWAVE model, which describes the decay of a vector particle to a
vector and a scalar~\cite{simulation1,simulation2}, and the fraction
of $S$-wave and $D$-wave is fixed according to the Belle
measurement~\cite{VVS}. The \edts\ process is generated via $D$-wave
with $D_{s2}^{*}(2573)^{-}$ decaying to $\bar{D}^0K^-$ via $D$-wave.
The $D_s^+\to K^-K^+\pi^+$ decay is simulated with the D\_Dalitz 
model~\cite{simulation1,simulation2}, and the $K_S^0\to
\pi^+\pi^-$ and $D_s^+\to K_S^0K^+$ decays are simulated with a phase
space model~\cite{simulation1,simulation2}.  Beam energy spread and
ISR are considered with the generator {\sc kkmc}~\cite{kkmc1,kkmc2}.

In the inclusive measurement, a $D_s^+$ is reconstructed with the
decay of \kkpi. The selection criteria for charged tracks are
described in Ref.~\cite{evt_sel}.  The tracks used to reconstruct
$D_s^+$ are required to originate from a common vertex, and the
$\chi^2$ of the vertex fit  ($\chi^2_{\rm VF}$)~\cite{svt} is required to
satisfy $\chi_{\rm VF}^2<100$.  Only decays containing the intermediate
states $\phi$ or $\bar{K}^{*0}$ in $D_s^+\rightarrow K^-K^+\pi^+$ are
used to select $D_s^+$ candidates.  The invariant masses of $K^+K^-$
($M(K^+K^-)$) or $K^-\pi^+$ ($M(K^-\pi^+)$) are required to satisfy
$1.004<M(K^+K^-)<1.034$~GeV/$c^2$ with a helicity angle of $K^+$ in
the $K^+K^-$ helicity frame satisfying
$|\cos\theta_{K^+/K^+K^-}|>0.4$, or
$0.832<M(K^-\pi^+)<0.928$~GeV/$c^2$ with
$|\cos\theta_{\pi^+/K^-\pi^+}|>0.52$.  The invariant mass of
$K^-K^+\pi^+$ ($M(K^-K^+\pi^+)$) is constrained to the known $D_s^+$
mass $m_{D_s^+}$~\cite{PDG} using a one-constraint
kinematic fit to improve the resolution of the $D_s^+$ recoiling
mass, $RM(D_s^+)$.

\begin{figure}[htbp]
    \centering
    \includegraphics[width=0.23\textwidth]{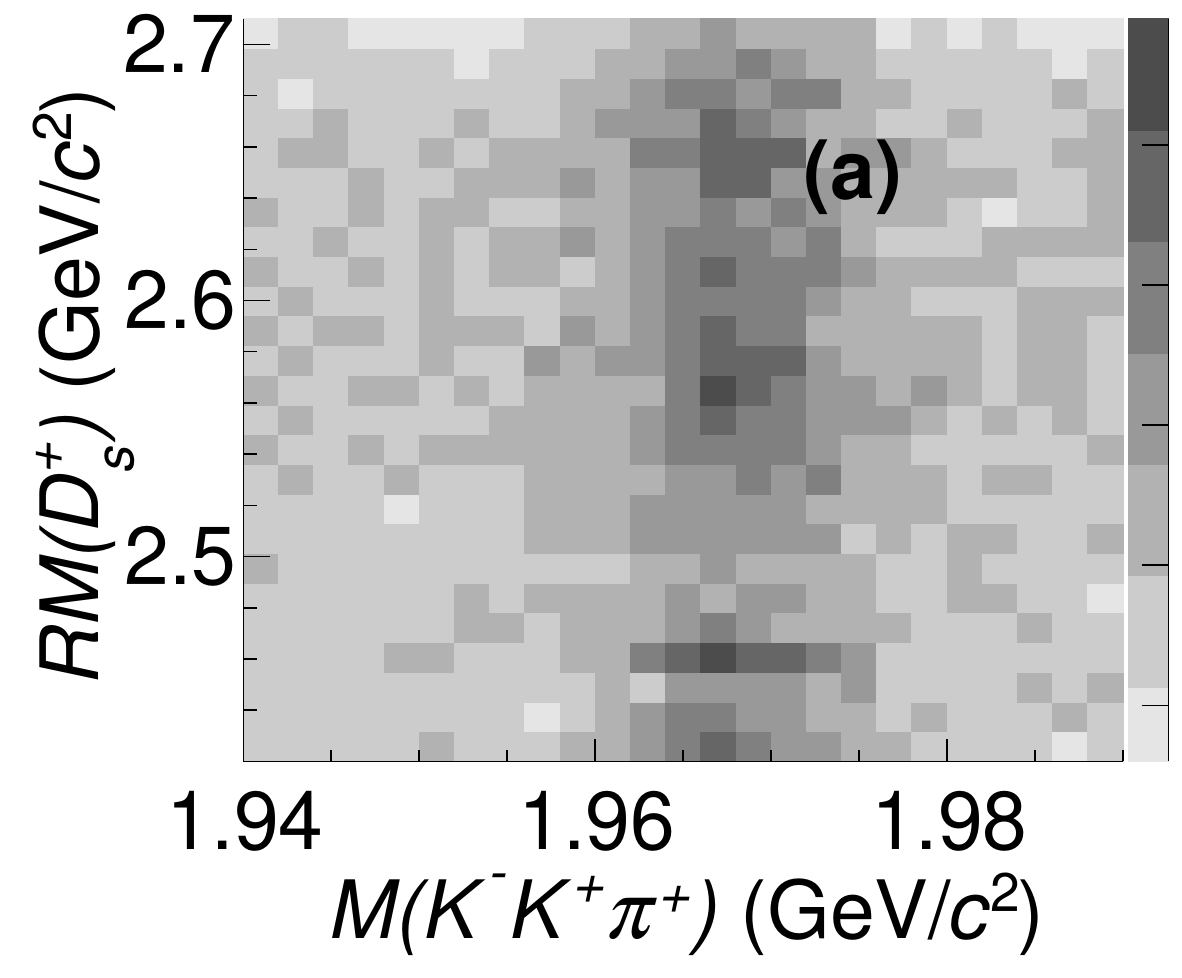}
    \includegraphics[width=0.23\textwidth]{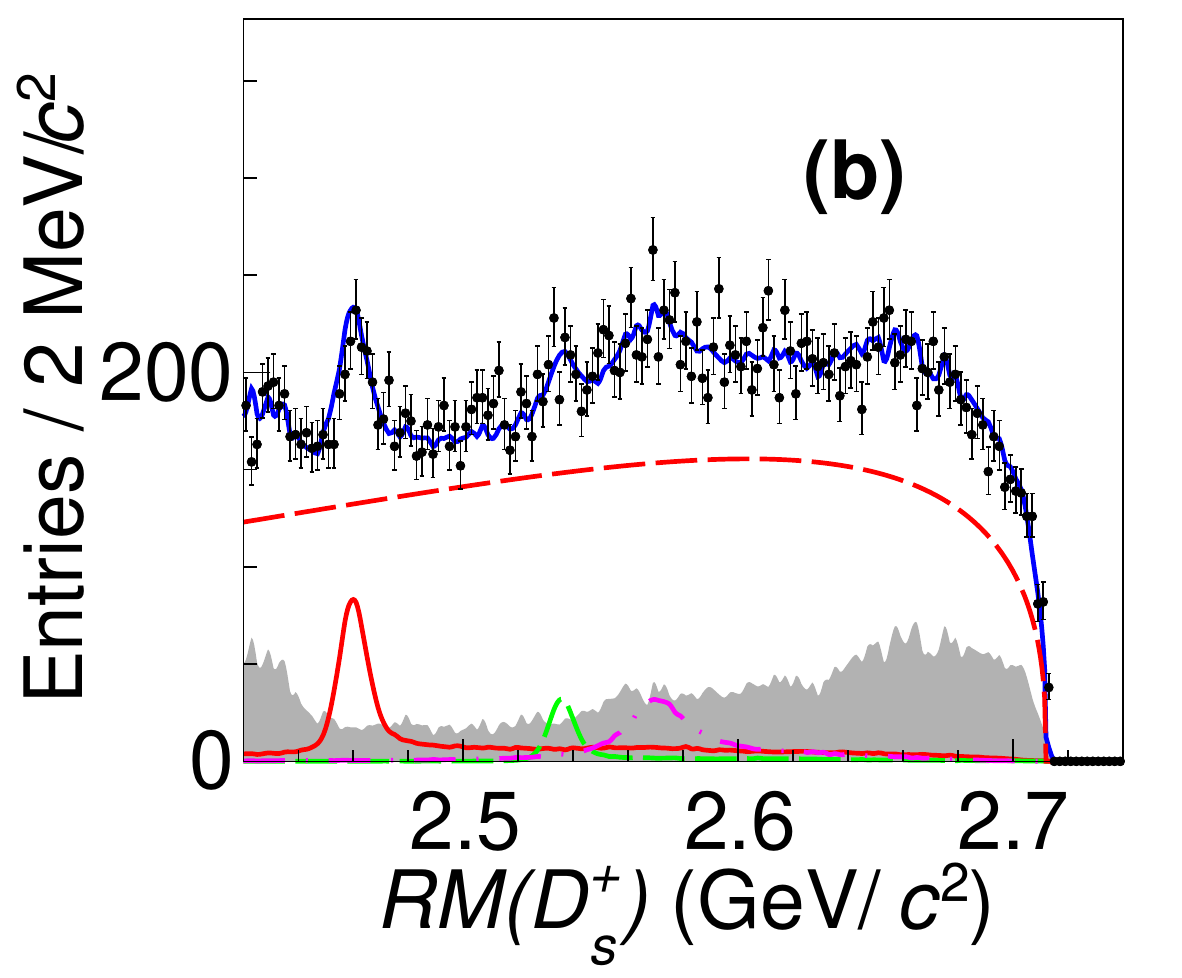}
\caption{ Distribution of (a) $RM(D_s^+)$ versus $M(K^-K^+\pi^+)$ from
  data and (b) projection of the 2D fit in $RM(D_s^+)$ in the
  inclusive analysis at $\sqrt{s}=4.680$~GeV.  Here, the dots with
  error bars are data, the gray histogram is background from processes
  involving an excited $D_s$ or $D$ meson, the red dashed line is an
    ARGUS function~\cite{argus}, the blue solid line is the total fit, and the red
  solid, green dashed, and purple dash-dotted lines are MC shapes of
  $D_{s1}(2460)^-$, $D_{s1}(2536)^-$, and $D_{s2}^{*}(2573)^-$
  signals, respectively.  }
    \label{pic:inc_xs}
\end{figure}

The yields of $D_{s1}(2536)^-$ and $D_{s2}^*(2573)^-$ events are
determined by a two-dimensional (2D) extended unbinned likelihood fit
to $M(K^-K^+\pi^+)$ versus $RM(D_s^+)$.  Distributions of $RM(D_s^+)$
versus $M(K^-K^+\pi^+)$ from data and the projection of the 2D fit in
$RM(D_s^+)$ at $\sqrt{s}=4.680$~GeV are shown in
Figs.~\ref{pic:inc_xs}(a) and~\ref{pic:inc_xs}(b), respectively. The
details of the fit methods in inclusive and exclusive measurements and
numerical results of the cross section calculation are described in
the supplemental material.  The cross sections are calculated with

\begin{equation}
    \sigma_{i,j}^{{\rm inc}}=\FS{N^{{\rm inc}}_{i,j}}{\FS{1}{|1-\Pi|^2}(1+\delta)_{i,j}\epsilon^{{\rm inc}}_{i,j}\mathcal{B}_{K^-K^+\pi^+}\mathcal{L}},
\end{equation}
where $\mathcal{B}_{K^-K^+\pi^+}$ is the branching fraction of
$\dskkp$ \cite{PDG}, $N^{{\rm inc}}_{i,j}$ is the number of signal
events obtained from the 2D fit, $(1+\delta)_{i,j}$ is the ISR correction
factor obtained from MC simulation, and $\epsilon^{{\rm inc}}_{i,j}$
is the detection efficiency for \edos\ ($i=1$) or \edts\ ($i=2$) in the
inclusive cross section measurement at the $j^{{\rm th}}$ $\sqrt{s}$;
$\FS{1}{|1-\Pi|^2}$ and $\mathcal{L}$ are the vacuum polarization factor
and integrated luminosity at the corresponding $\sqrt{s}$,
respectively.

In the exclusive measurement, a $D_s^+$ is reconstructed with the
decay of \kkpi\ or \ksk\ and a $K^-$ is selected from the charged
tracks not forming the $D_s^+$.  The selection criteria for $K_S^0$
are described in Refs.~\cite{evt_sel, svt}. The tracks used to
reconstruct $D_s^+$, including the virtual track of $K_S^0$ from a
secondary vertex fit~\cite{svt}, are also required to originate from a
common vertex with $\chi_{\rm VF}^2<100$.  In addition to the selection
criteria used in the inclusive analysis, the invariant mass of
$K^-K^+\pi^+$ or $K_S^0K^+$ ($M(K_S^0K^+)$) must satisfy
$|M(K^-K^+\pi^+/K_S^0K^+)-m_{D_s^+}|<8$~MeV/$c^2$.  To select $\dtd$
and $\dod$, the recoiling mass of $D_s^+K^-$ ($RM(D_s^+K^-)$) must
satisfy the following requirements: for the $e^+e^-\to
D_s^+D_{s1}(2536)^-$ process, $|RM(D_s^+K^-)-m_{\bar{D}^{*0}}|$ should
be less than $9$~MeV/$c^2$ for $D_s^+\to K^-K^+\pi^+$ and
$7$~MeV/$c^2$ for $D_s^+\to K_S^0K^+$; for the $e^+e^-\to
D_s^+D^*_{s2}(2573)^-$ process, $|RM(D_s^+K^-)-m_{\bar{D}^0}|$ should
be less than $11$~MeV/$c^2$ for $D_s^+\to K^-K^+\pi^+$ and
$9$~MeV/$c^2$ for $D_s^+\to K_S^0K^+$.  Here,
$m_{\bar{D}^{*0}}$ and
$m_{\bar{D}^0}$ are the known $\bar{D}^{*0}$ and $\bar{D}^0$ masses, respectively~\cite{PDG}.  For
the selected entries, $M(K^-K^+\pi^+/K_S^0K^+)$ is constrained to
$m_{D_s^+}$, $RM(D_s^+K^-)$ is constrained to $m_{\bar{D}^0}$ or
$m_{\bar{D}^{*0}}$, and the total four-momentum is constrained to that
of the initial $e^+e^-$ system via a kinematic fit.

\begin{figure}[htbp]
    \centering
    \includegraphics[width=0.22\textwidth]{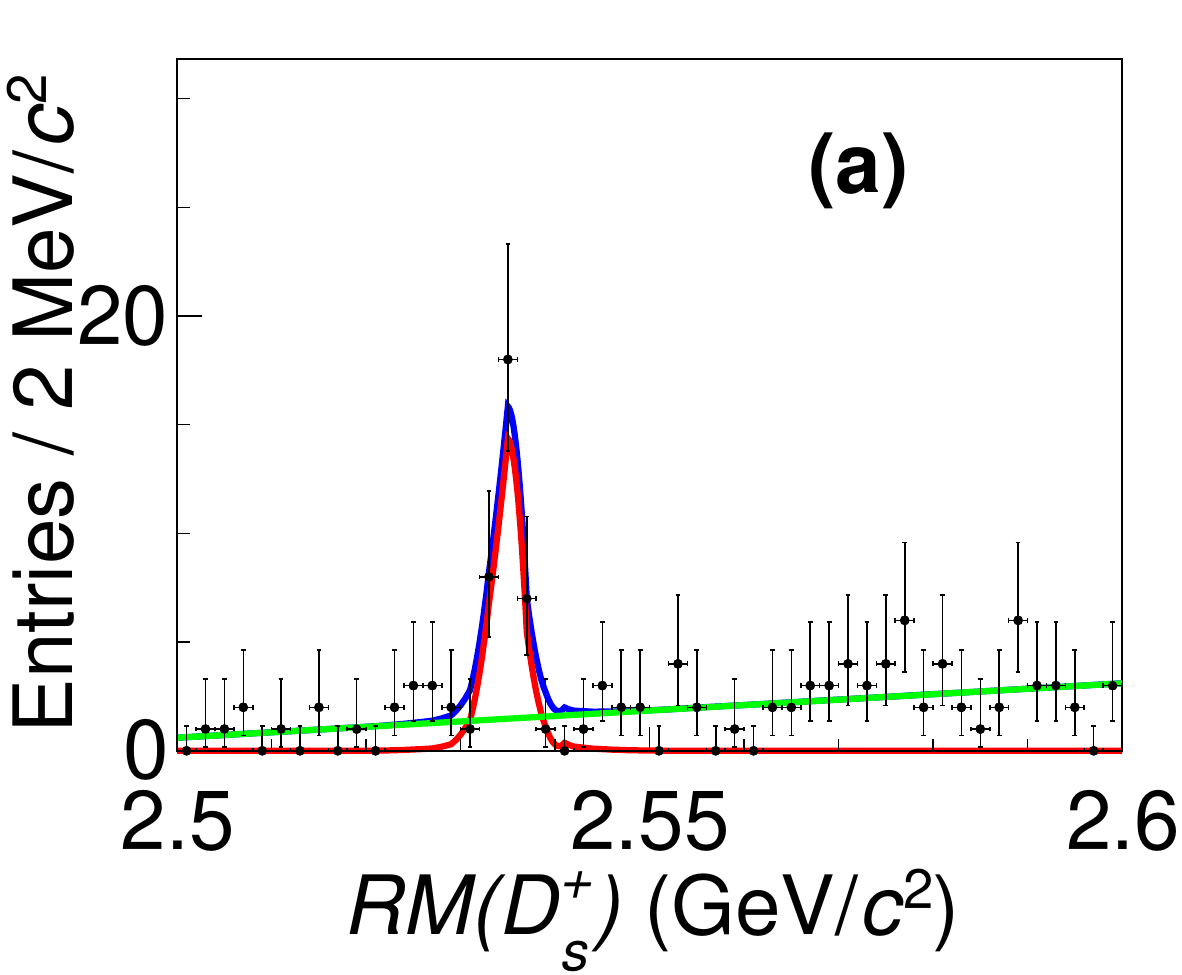}
    \includegraphics[width=0.22\textwidth]{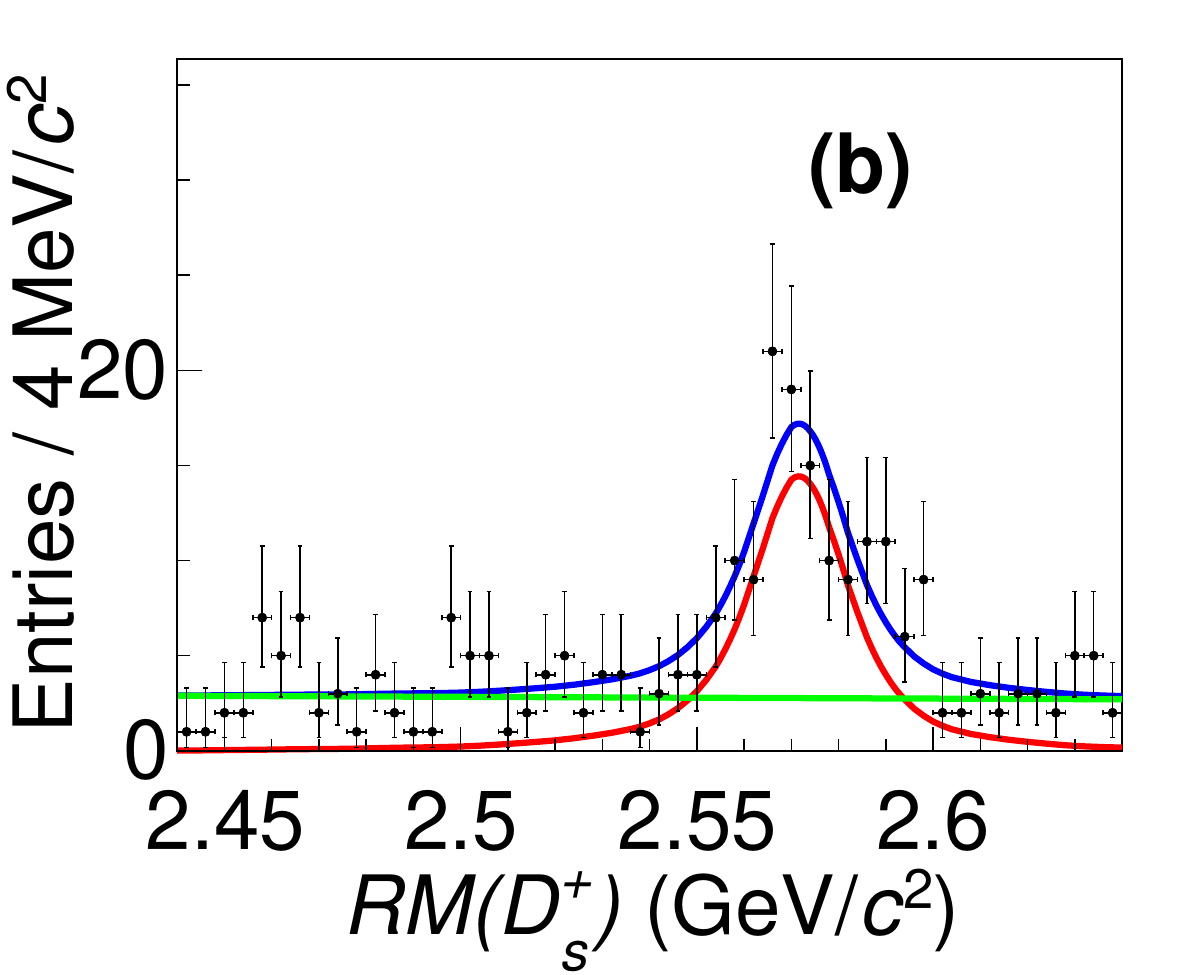}
\caption{ Fit results of $RM(D_s^+)$ for (a) $D_{s1}(2536)^-$  and
  (b) $D^*_{s2}(2573)^-$  in the exclusive analysis at
  $\sqrt{s}=4.680$~GeV. Here, the dots with error bars are data, the
  blue, red, and green solid lines are the total fit, signal shape,
  and background shape, respectively.  }
    \label{pic:ex_fit}
\end{figure}

\begin{figure}[htbp]
    \centering
    \includegraphics[width=0.22\textwidth]{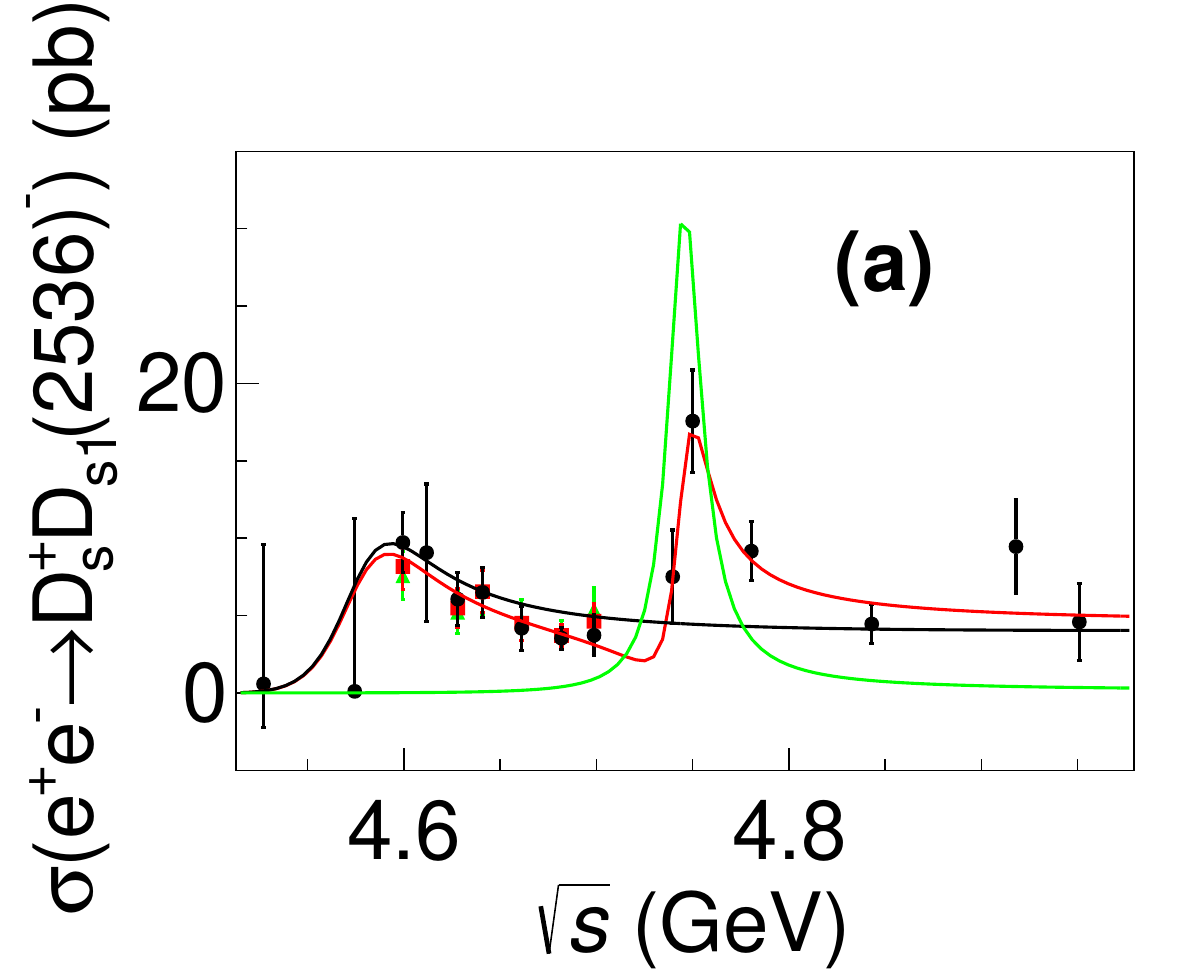}
    \includegraphics[width=0.22\textwidth]{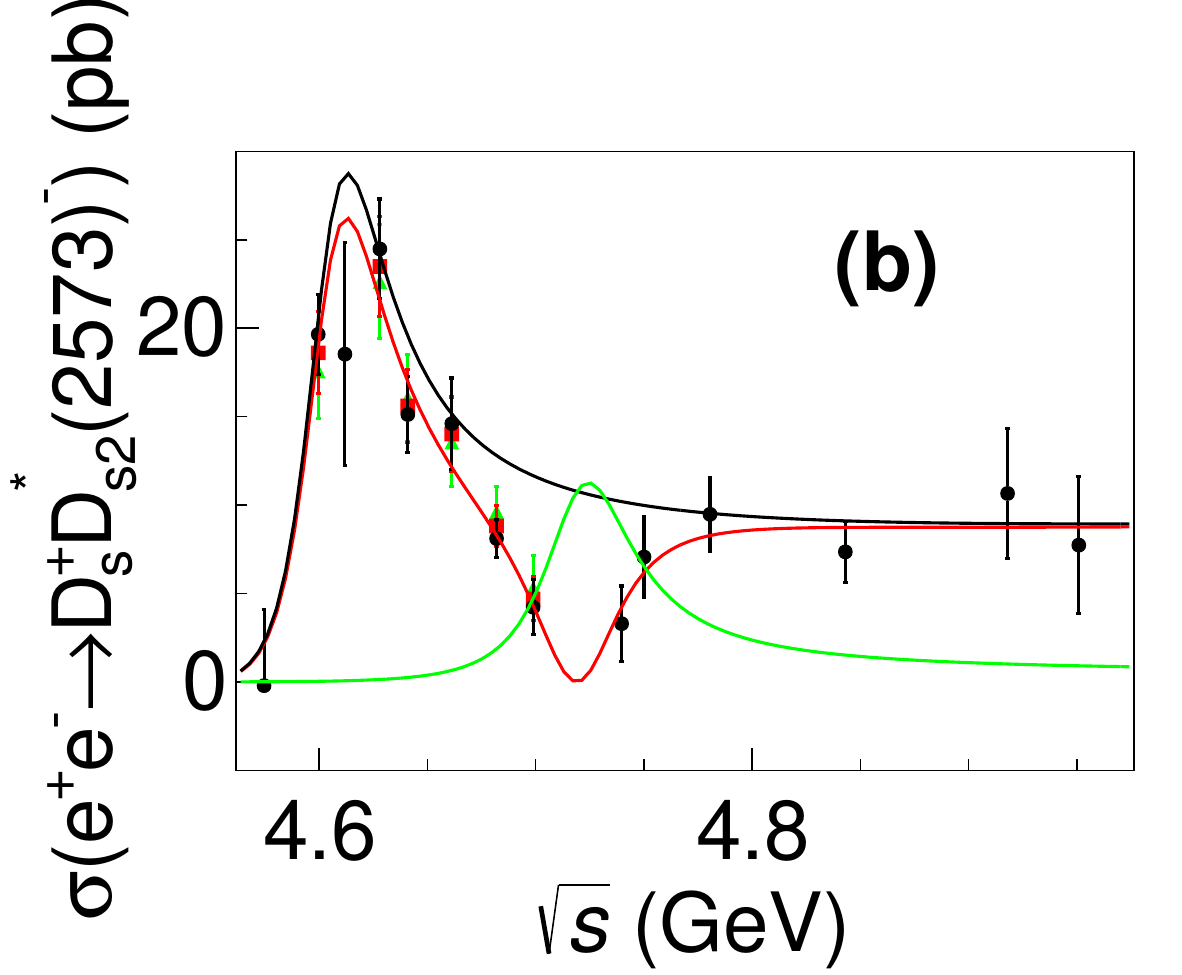}
\caption{ Cross sections of (a) \edos\ with $\dod$ and (b) \edts\ with
  $\dtd$. The black dots, red squares, and green triangles with
  error bars are measured exclusive cross sections, inclusive cross
  section from likelihood fit multiplied by the absolute branching
  fraction, and measured inclusive cross section multiplied by the
  absolute branching fraction, respectively. The red, black, and green
  solid lines are results of total fit, $BW_0$, and $BW_1$,
  respectively. The uncertainties are statistical only.  }
    \label{pic:ex_xs_fit}
\end{figure}

For data samples with $\sqrt{s}\geq4.6$~GeV, the yields of
$D_{s1}(2536)^-$ and $D^*_{s2}(2573)^-$ events are determined by 
extended unbinned likelihood fits to the corresponding $RM(D_s^+)$ distributions,
while for data samples with $\sqrt{s}<4.6$~GeV, due to the low number
of events, the counting method described in Refs.~\cite{count1,
  count2} is used. The fit results of $RM(D_s^+)$ for $D_{s1}(2536)^-$
and $D^{*}_{s2}(2573)^-$ at $\sqrt{s}=4.680$~GeV are shown in
Figs.~\ref{pic:ex_fit}(a) and~\ref{pic:ex_fit}(b), respectively.  The
cross sections are calculated with
\begin{equation}
    \label{eq_ex_xs}
    \sigma^{{\rm exc}}_{i,j}=\FS{ N^{{\rm exc}}_{i,j}}{\FS{1}{|1-\Pi|^2}(1+\delta)_{i,j}(\epsilon\mathcal{B})_{i,j} \mathcal{L}},
\end{equation}
where $N^{{\rm exc}}_{i,j}$ is the number of signal events obtained
from the fit and $(\epsilon\mathcal{B})_{i,j}=(\epsilon^{{\rm
    exc}}_{K^-K^+\pi^+, i,j} \mathcal{B}_{K^-K^+\pi^+}+\epsilon^{{\rm
    exc}}_{K_S^0K^+, i,j}\mathcal{B}_{K_S^0K^+})$. Here,
$\mathcal{B}_{K_S^0K^+}=\mathcal{B}(\dsksk)\mathcal{B}(K_S^0\to
\pi^+\pi^-)$ \cite{PDG} is the product of the branching fractions of
$\dsksk$ and $K_S^0\to \pi^+\pi^-$, $\epsilon^{{\rm
    exc}}_{K^-K^+\pi^+, i,j}$ and $\epsilon^{{\rm exc}}_{K_S^0K^+,
  i,j}$ are the detection efficiencies for the signal processes with
$\dskkp$ and \ksk, respectively.  The measured cross sections of
\edos\ and \edts\ with the inclusive and exclusive methods are shown
in Figs.~\ref{pic:ex_xs_fit}(a) and~\ref{pic:ex_xs_fit}(b),
respectively.

Using the data at the six energy points with both inclusive and
exclusive cross sections measured, we determine the absolute branching
fractions of the $\dod$ and $\dtd$ with a likelihood fit
that maximizes the likelihood function, 
\begin{equation}
    \label{nll_func}
    \begin{aligned}
        & L_{i}(\sigma_{i,j}^{{\rm inc}},\delta_{i,j}^{{\rm inc}},\sigma_{i,j}^{{\rm exc}},\delta_{i,j}^{{\rm exc}};\sigma_{i,j},\mathcal{B}_{i})=\\
        & \prod_{j=1}^6{L_{i,j}^{{\rm inc}}(\sigma^{{\rm inc}}_{i,j},\delta^{{\rm inc}}_{i,j};\sigma_{i,j})L_{i,j}^{{\rm exc}}(\sigma^{{\rm exc}}_{i,j},\delta^{{\rm exc}}_{i,j};\sigma_{i,j},\mathcal{B}_{i})},  
    \end{aligned}
\end{equation}
where $\delta_{i,j}^{{\rm inc}}$ and $\delta_{i,j}^{{\rm exc}}$ are
the statistical uncertainties of the measured inclusive and exclusive
cross sections, respectively; $\sigma_{i,j}$ is the actual cross
section of \edos\ or \edts; and $\mathcal{B}_{i}$ is the absolute
branching fraction of $\dod$ ($i=1$) or $\dtd$ ($i=2$).  Since the
significances for \edos\ (\edts) at $\sqrt{s}=4.66$ (4.66 and 4.7)~GeV
in both inclusive and exclusive measurements are less than 5$\sigma$,
$L_{i,j}^{{\rm inc, exc}}$ at that energy point is a normalized
likelihood as a function of $\sigma_{i,j}^{{\rm inc,
    exc}}$ which is obtained from the signal
yield fits.  The likelihood $L^{{\rm inc, exc}}_{i,j}$ for the other
samples with sufficiently high statistics is approximated as a
Gaussian function, and details are described in the supplemental material.
Figures~\ref{pic:br}(a) and~\ref{pic:br}(b) show the fit results of
the absolute branching fractions, which are $(35.9\pm 4.8)\%$ and
$(37.4\pm 3.1)\%$ for $\mathcal{B}(\dod)$ and $\mathcal{B}(\dtd)$,
respectively.

\begin{figure}[htbp]
    \centering
    \includegraphics[width=0.22\textwidth]{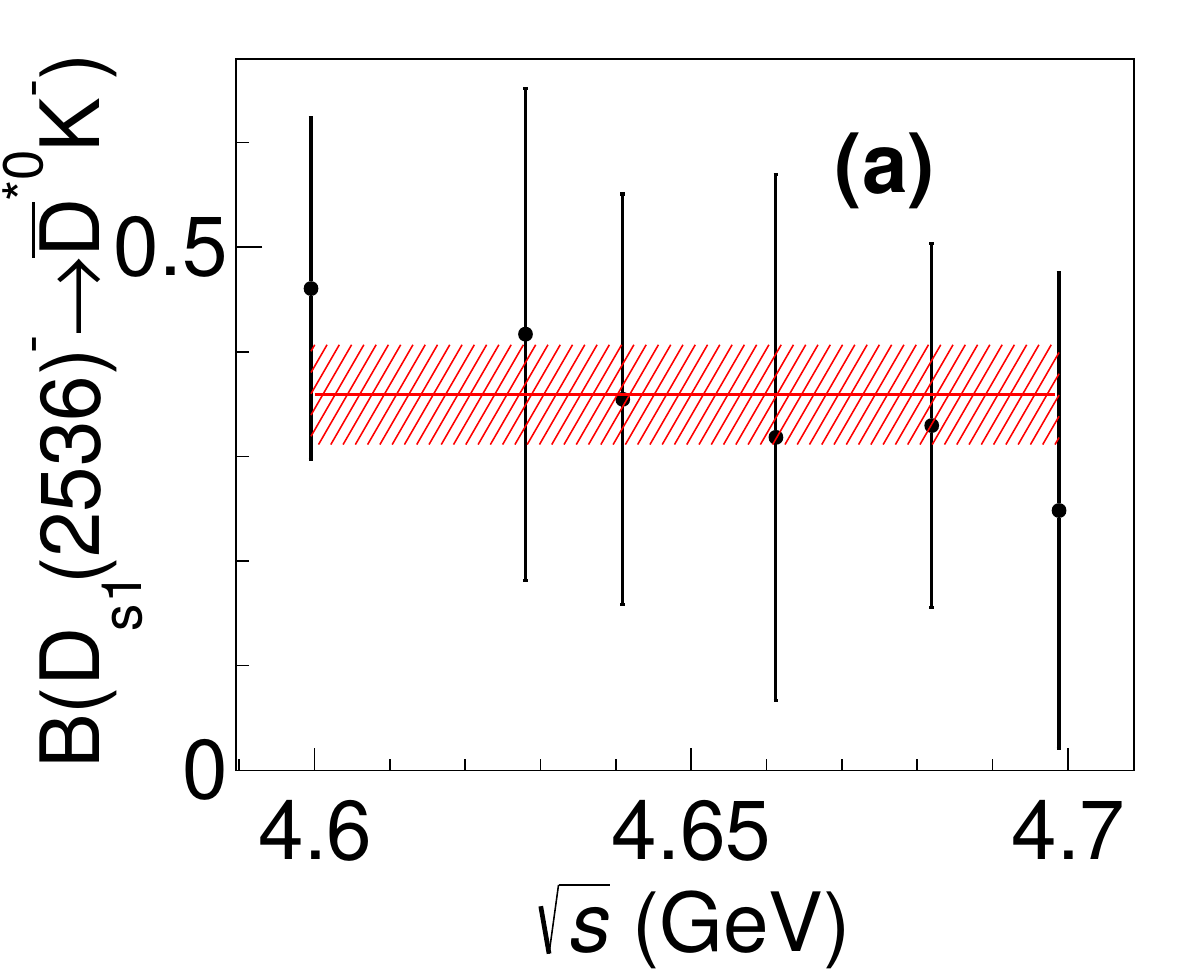}
    \includegraphics[width=0.22\textwidth]{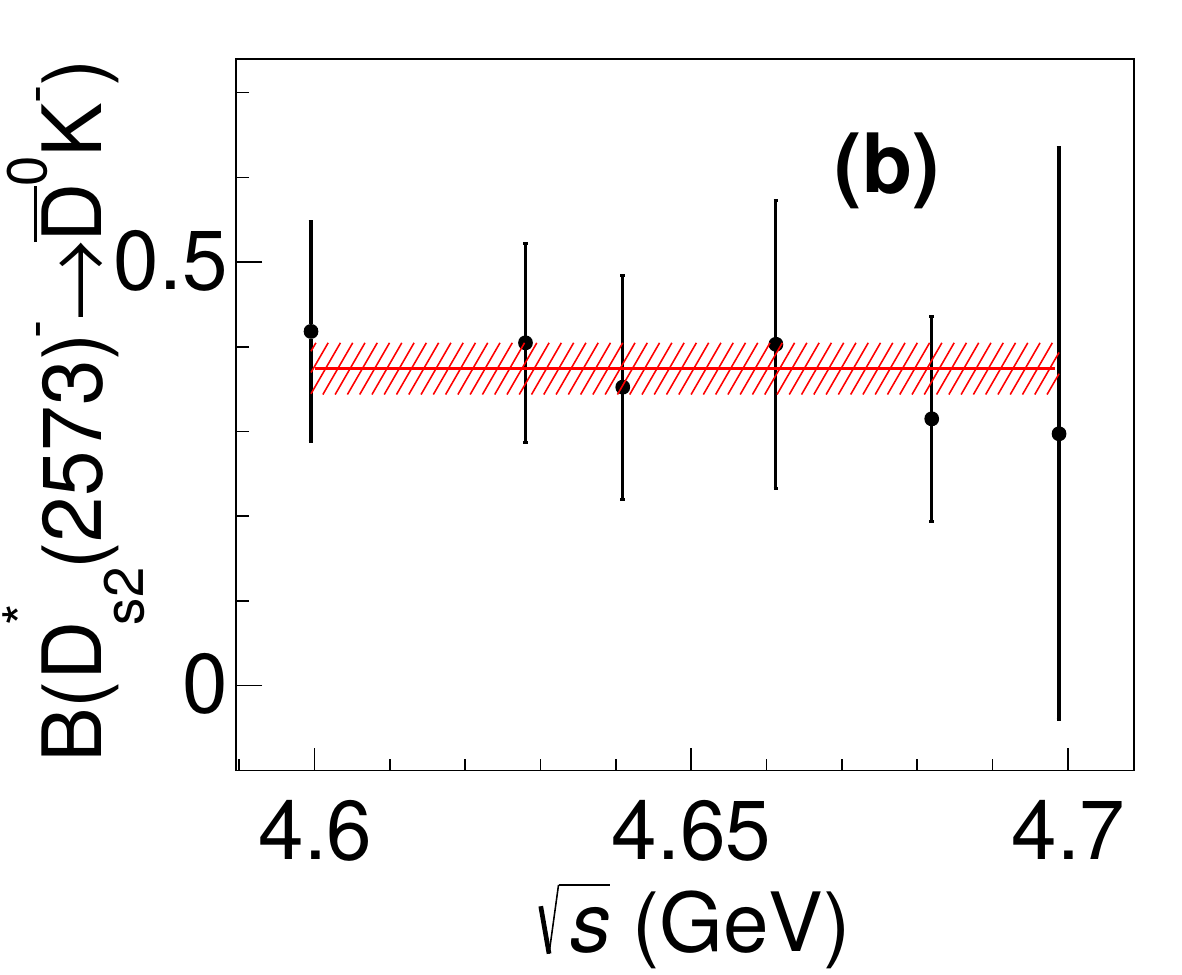}
\caption{ The absolute branching fractions of (a) $\dod$ and (b)
  $\dtd$. The black dots with error bars are absolute branching
  fractions calculated at each $\sqrt{s}$, where
  $\mathcal{B}_{i,j}=\sigma^{{\rm exc}}_{i,j}/\sigma^{{\rm
      inc}}_{i,j}$. The red lines represent results calculated by the
  maximum likelihood fit. The uncertainties are statistical only and
  are shown with the red shaded bands.  }
    \label{pic:br}
\end{figure}

To study the resonance structures in the \edos\ and \edts\ processes,
least-$\chi^2$ fits to the measured cross sections are performed. The
cross sections are described with the coherent sum of two constant-width
Breit-Wigner ($BW$) functions.  The fit results are shown in
Figs.~\ref{pic:ex_xs_fit}(a) and~\ref{pic:ex_xs_fit}(b) with
$\chi^2/{\rm ndf}=4.0/8$ and $6.2/7$, respectively, where ${\rm
ndf}$ is the number of degrees of freedom, and the fit
  details are described in the supplemental material.  By comparing
  $\Delta \chi^2$ of the fits with and without the corresponding
  component and accounting for $\Delta{\rm ndf}$, the significance is
  determined. The statistical significances of the first and second
  resonance structures are $7.2\sigma$ and $4.3\sigma$, respectively,
  in \edos, and $15\sigma$ and $2.7\sigma$, respectively, in \edts.
  In both processes, the first resonance structure is around 4.6 GeV
  with a width of 50 MeV. In \edos, the second one is around 4.75 GeV
  with a width of 25 MeV, and in \edts, around 4.72 GeV with a width
  of 50 MeV. Continuum contributions are also tested, but the
  significances are less than 1$\sigma$ in both processes.

The systematic uncertainties for the measurements of absolute
branching fractions related to fits, including signal and background
descriptions and fit ranges in the fits of inclusive and exclusive
analyses, are described in the supplemental material. The other systematic
uncertainties are introduced below.

The systematic uncertainties from the mass window requirement of
$M(D_s^+)$ ($RM(D_s^+K^-)$) are estimated by comparing the efficiency
difference between data and MC simulation~\cite{eff_comp} as 3.4\% and
5.5\% (4.3\% and 4.3\%), for $\dod$ and $\dtd$, respectively.

The systematic uncertainties from tracking (particle identification,
PID) efficiencies for $K^\pm$ and $\pi^+$ from $D_s^+$ are taken as
0.5\% (0.5\%) and 0.2\% (0.4\%), respectively~\cite{DsstDsst}. The
systematic uncertainty from $K_S^0$ reconstruction is assigned as
2.3\%~\cite{ks_rec}. Most of these uncertainties cancel in the $D_s^+$
reconstruction as they appear in both inclusive and exclusive
processes. Only those uncertainties not common between the two are
considered, and the systematic uncertainties from $D_s^+\rightarrow
K^-K^+\pi^+$ and $D_s^+\rightarrow K_S^0K^+$ are added according to
their branching fractions.  Since the momentum of the bachelor $K^-$
that does not come from $D_s^+$ decays in the exclusive analysis is
very low, the systematic uncertainties of this $K^-$ are estimated
with a control sample of $J/\psi\rightarrow
pK^{-}\Lambda$~\cite{trk_pid} as 1.2\% and 0.0\% for $\dod$ and
$\dtd$, respectively.

The uncertainties of $\mathcal{B}(\dskkp)$ and $\mathcal{B}(\dsksk)$
are 1.9\% and 2.4\%~\cite{PDG}, respectively.  The systematic
uncertainty from $\mathcal{B}(\dskkp)$ cancels out in the calculation
of the absolute branching fractions, but does not cancel in the
exclusive cross section measurements.

The fractions of the $S$-wave and $D$-wave of the $D_{s1}(2536)^-\to
\bar{D}^{*0}K^-$ decay are changed by one standard deviation, and the
systematic uncertainty is estimated by the maximum change at
$\sqrt{s}=4.680$~GeV on the exclusive cross section as 0.2\%.

The total systematic uncertainties are 9.7\% and 12.4\% for
the two processes, respectively, by assuming all sources to be
independent and summing them in quadrature.

Most systematic uncertainty estimations for the exclusive cross
section measurements are the same as those described for the absolute
branching fraction measurements, including the mass window
requirements, $\mathcal{B}(\dskkp)$ and $\mathcal{B}(\dsksk)$, the
fraction of the $S$-wave and $D$-wave in the $D_{s1}(2536)^-\to
\bar{D}^{*0}K^-$ decay, and tracking and PID efficiencies, where 1.9\% is
assigned for tracks from $D_s^+$ for both processes. Systematic
uncertainties related to the fit, including the fit range and
background shape, are described in the supplemental
material. Additional sources of systematic uncertainties unique to the
exclusive cross section measurement are described below.

The angular distribution of $e^+e^-\to D_s^+D_{s1}(2536)^{-}$ is
described by $1+\alpha {\rm cos}^2\theta$ with the {\sc AngSam} model. To
estimate the systematic uncertainty from this model, $\alpha$ is
changed by one standard deviation and the maximum change at
$\sqrt{s}=$4.680 GeV is taken as the uncertainty of 3.3\%.  The ISR
correction factor and efficiency of the signal process depend on the
input cross section in {\sc kkmc}. We sample the input cross section
500 times at each $\sqrt{s}$ according to its statistical uncertainty,
and take the ratio of the standard deviation and the mean value of
$\epsilon(1+\delta)$ as the systematic uncertainty.  The uncertainty
from the luminosity measurement is 1\%~\cite{Lum1,XYZEcm2}.

The systematic uncertainties introduced above, as well as the total
ones are shown in Tables \ref{table:sys_ex_xs_DsDs1_1} and
\ref{table:sys_ex_xs_DsDs2_1}. Tables with all systematic
uncertainties are provided in the supplemental material.  The
systematic uncertainties of the data sample at $\sqrt{s}=4.600$~GeV
are assigned to those of the data samples at $\sqrt{s}=4.530$ and
4.575~GeV because of low statistics.

\begin{table*}[htp]
	\centering
	\caption{Systematic uncertainties (\%) in the cross sections for \edos. ``..." represents systematic uncertainties related to fit or common among data samples.}
	\label{table:sys_ex_xs_DsDs1_1}
		\begin{tabular}{cccccccccccccc}
			\hline\hline
			$\sqrt{s}$ (GeV) & 4.600 & 4.610 & 4.620 & 4.640 & 4.660 & 4.680 & 4.700 & 4.740 & 4.750 & 4.780 & 4.840 & 4.914 & 4.946 \\
			\hline
			Tracking \& PID ($K^{\pm}$ not from $D_s^+$) & 1.7  & 1.5  & 1.4  & 1.6  & 1.3  & 1.2  & 1.1  & 0.9  & 0.9  & 0.8  & 0.8  & 0.8  & 0.8  \\
			ISR                              & 1.4  & 3.5  & 3.6  & 1.9  & 2.2  & 1.7  & 2.0  & 3.2  & 1.1  & 1.0  & 0.9  & 0.4  & 2.3  \\
			...                       & ...  & ...  & ...  & ...  & ...  & ...  & ...  & ...  & ...  & ...  & ...  & ...  & ...  \\
			Total                            & 9.2  & 9.2  & 12.5 & 10.1 & 10.0 & 7.9  & 8.0  & 8.3  & 11.2 & 9.8  & 10.3 & 11.2 & 15.6 \\
			\hline\hline
		\end{tabular}
\end{table*}

\begin{table*}[htp]
	\centering
	\caption{Relative systematic uncertainties (\%) in the cross section for \edts. Definition of ``..." is the same as in Table \ref{table:sys_ex_xs_DsDs1_1}.}
	\label{table:sys_ex_xs_DsDs2_1}
		\begin{tabular}{cccccccccccccc}
			\hline\hline
			$\sqrt{s}$ (GeV) & 4.600 & 4.610 & 4.620 & 4.640 & 4.660 & 4.680 & 4.700 & 4.740 & 4.750 & 4.780 & 4.840 & 4.914 & 4.946 \\
			\hline
			Tracking \& PID ($K^{\pm}$ not from $D_s^+$) & 0.1  & 0.1  & 0.0  & 0.0  & 0.0  & 0.1  & 0.1  & 0.1  & 0.0  & 0.1  & 0.0  & 0.1  & 0.1  \\
			ISR                              & 1.0  & 2.7  & 1.1  & 1.3  & 0.8  & 1.2  & 2.9  & 5.3  & 2.5  & 0.7  & 0.6  & 0.6  & 1.6  \\
			...                              & ...  & ...  & ...  & ... & ...  & ...  & ...  & ...  & ...  & ...  & ... & ... & ... \\
			Total                            & 8.5  & 10.6 & 9.7  & 15.6 & 9.7  & 11.9 & 11.0 & 9.9  & 9.3  & 10.7 & 15.2 & 47.0 & 27.0 \\
			\hline\hline
		\end{tabular}
\end{table*}

In summary, we measure for the first time the absolute branching
fractions of $\dod$ and $\dtd$ as $(35.9\pm 4.8\pm 3.5)\%$ and
$(37.4\pm3.1\pm4.6)\%$, respectively, where the first uncertainties
are statistical and the second systematic. Assuming isospin symmetry
and neglecting the phase space differences, we obtain
$\mathcal{B}(D_{s1}(2536)^{-}\rightarrow(\bar{D}^{*}\bar{K})^{-})=(71.8\pm
9.6\pm 7.0)\%$ and
$\mathcal{B}(D^{*}_{s2}(2573)^{-}\rightarrow(\bar{D}\bar{K})^{-})=(74.8\pm
6.2\pm 9.2)\%$.
$\mathcal{B}(D_{s1}(2536)^{-}\rightarrow(\bar{D}^{*}\bar{K})^{-})$
($\mathcal{B}(D^{*}_{s2}(2573)^{-}\rightarrow(\bar{D}\bar{K})^{-})$)
is more than two (one) standard deviations from the prediction of
Refs.~\cite{frac_Dsj_ex, frac_Dsj_ex2}, about 100\% (90\%), if
$D_{s1}(2536)$ ($D_{s2}^{*}(2573)$) is predominantely a bare $c\bar{s}$
meson.  Our measurements indicate that non-$c\bar{s}$ components may
exist in the $D_{s1}(2536)$ and $D_{s2}^*(2573)$ wave functions.  The
exclusive cross sections of \edos\ with $\dod$ and \edts\ with $\dtd$
are also reported in this Letter.  A resonant structure at around
4.6~GeV is observed for the first time in \edts, which is consistent
with the evidence for the $Y(4620)$ with the same final state reported
by the Belle collaboration~\cite{DsDs2573Belle}. A clear enhancement
at around 4.6~GeV is also observed in \edos, which could be the
$Y(4626)$ state observed by the Belle
collaboration~\cite{DsDs2536Belle} in the same final state. Our data
may indicate that the same state at around 4.6~GeV decays into both
$D_s^+D_{s1}(2536)^{-}$ and $D_s^+D_{s2}^*(2573)^{-}$ final states.
Evidence for a structure at around 4.75~GeV is observed, which may be
the $Y(4710)$ or $Y(4790)$ reported earlier by the BESIII
experiment~\cite{Y4710KKJpsiBESIII,Y4790KKJpsiBESIII}.


\textbf{Acknowledgment}

The BESIII Collaboration thanks the staff of BEPCII and the IHEP computing center for their strong support. This work is supported in part by National Key R\&D Program of China under Contracts Nos. 2020YFA0406300, 2020YFA0406400, 2023YFA1606000; National Natural Science Foundation of China (NSFC) under Contracts Nos. 11635010, 11735014, 11935015, 11935016, 11935018, 12025502, 12035009, 12035013, 12061131003, 12192260, 12192261, 12192262, 12192263, 12192264, 12192265, 12221005, 12225509, 12235017, 12361141819; the Chinese Academy of Sciences (CAS) Large-Scale Scientific Facility Program; the CAS Center for Excellence in Particle Physics (CCEPP); Joint Large-Scale Scientific Facility Funds of the NSFC and CAS under Contract No. U1832207; 100 Talents Program of CAS; The Institute of Nuclear and Particle Physics (INPAC) and Shanghai Key Laboratory for Particle Physics and Cosmology; German Research Foundation DFG under Contracts Nos. 455635585, FOR5327, GRK 2149; Istituto Nazionale di Fisica Nucleare, Italy; Ministry of Development of Turkey under Contract No. DPT2006K-120470; National Research Foundation of Korea under Contract No. NRF-2022R1A2C1092335; National Science and Technology fund of Mongolia; National Science Research and Innovation Fund (NSRF) via the Program Management Unit for Human Resources \& Institutional Development, Research and Innovation of Thailand under Contract No. B16F640076; Polish National Science Centre under Contract No. 2019/35/O/ST2/02907; The Swedish Research Council; U. S. Department of Energy under Contract No. DE-FG02-05ER41374.

\end{document}



\begin{center}
   {{\bf \boldmath Supplemental Material for ``Study of the decay and production properties of $D_{s1}(2536)$ and $D_{s2}^{*}(2573)$"}}
\end{center}
\maketitle
\bigskip

\section{Data Samples}
This analysis is performed based on the data samples collected with the BESIII detector operating at BEPCII. The center-of-mass energies ($\sqrt{s}$) and the corresponding integrated luminosities of these samples ($\mathcal L$) are  listed in Table \ref{table:dataset}. 

	\begin{table*}[htp]
		\centering
		\caption{The nominal $\sqrt{s}$ (mentioned in the main
                  body of Letter), measured $\sqrt{s}$, and integrated
                  luminosity of data sample ($\mathcal L$) used in
                  this Letter~\cite{XYZEcm1, XYZEcm2, Lum1}. For the
                  measured $\sqrt{s}$ and integrated luminosity, the
                  first uncertainty is statistical and the second
                  systematic.}
			\begin{tabular}{ccc}
				\hline\hline
				$\sqrt{s}^{{\rm nominal}}$ (GeV) & Measured $\sqrt{s}$ (MeV) & $\mathcal L$ ($\rm{pb}^{-1}$) \\
				\hline
				4.530  & 4527.14$\pm$0.11$\pm$0.72 	& 112.12$\pm$0.04$\pm$0.73   \\
				4.575  & 4574.50$\pm$0.18$\pm$0.70 	&  48.93$\pm$0.03$\pm$0.32   \\
				4.600  & 4599.53$\pm$0.07$\pm$0.74 	& 586.9$\pm$0.1$\pm$3.9   \\
				4.610  & 4611.86$\pm$0.12$\pm$0.32 	& 103.83$\pm$0.05$\pm$0.55   \\
				4.620  & 4628.00$\pm$0.06$\pm$0.32 	& 521.52$\pm$0.11$\pm$2.76	\\
				4.640  & 4640.91$\pm$0.06$\pm$0.38 	& 552.41$\pm$0.12$\pm$2.93	\\
				4.660  & 4661.24$\pm$0.06$\pm$0.29 	& 529.63$\pm$0.12$\pm$2.81	\\
				4.680  & 4681.92$\pm$0.08$\pm$0.29	    & 1669.31$\pm$0.21$\pm$8.85	\\
				4.700  & 4698.82$\pm$0.10$\pm$0.39 	& 536.45$\pm$0.12$\pm$2.84	\\
				4.740  & 4739.70$\pm$0.20$\pm$0.30 	& 164.27$\pm$0.07$\pm$0.87	\\
				4.750  & 4750.05$\pm$0.12$\pm$0.29 	& 367.21$\pm$0.10$\pm$1.95	\\
				4.780  & 4780.54$\pm$0.12$\pm$0.33 	& 512.78$\pm$0.12$\pm$2.72	\\
				4.840  & 4843.07$\pm$0.20$\pm$0.31 	& 527.29$\pm$0.12$\pm$2.79	\\
				4.914  & 4918.02$\pm$0.34$\pm$0.35	    & 208.11$\pm$0.02$\pm$1.10	\\
				4.946  & 4950.93$\pm$0.36$\pm$0.44 	& 160.37$\pm$0.07$\pm$0.85	\\
				\hline\hline
			\end{tabular}
		\label{table:dataset}
	\end{table*}

\section{Fit methods and numerical results in inclusive and exclusive measurements}
In the inclusive analysis, the  $D_{s1}(2536)^-$ and
$D_{s2}^*(2573)^-$ yields are determined by a two-dimensional (2D)
extended unbinned likelihood fit of $M(K^-K^+\pi^+)$ and
$RM(D_s^+)$. The probability distribution function (PDF) is defined
as
\begin{equation}
\begin{aligned}
    {\rm  PDF} = \sum S_{M(K^-K^+\pi^+)}\times S_{RM(D_s^+)} 
    + B_{M(K^-K^+\pi^+)}\times B_{RM(D_s^+)} + B_{{\rm Other\ B.K.G.}}.
\end{aligned}
\end{equation}
Here, $S_{M(K^-K^+\pi^+)}$ is the MC shape convolved with a Gaussian
function to account for the difference between data sample and MC
simulation. The MC shape is obtained from the $e^+e^-\to
D_s^+D_{sj}^-$ process, where $D_{sj}^-$ includes $D_{s1}(2460)^-$,
$D_{s1}(2536)^-$, and $D_{s2}^*(2573)^-$; the parameters of the
Gaussian function are fixed according to the other fit to
$M(K^-K^+\pi^+)$ using the MC shape obtained from the $e^+e^-\to
D_s^+D_{s1}(2536)^-$ process as the signal shape and a first order
Chebychev polynomial function as the background shape.
$B_{M(K^-K^+\pi^+)}$ is a first order Chebychev polynomial function
describing the $D_s^+$ background shape, $S_{RM(D_s^+)}$ is the MC
shape obtained from the $e^+e^-\to D_s^+D_{sj}^-$ process, and
$B_{RM(D_s^+)}$ is an ARGUS function describing the background shape
in $RM(D_s^+)$.  $B_{{\rm Other\ B.K.G.}}$ includes the MC shapes
obtained from the $\eedsst$, $e^+e^-\to D_s^{*+}D_{s0}^*(2317)^-$,
$e^+e^-\to D_s^{*+}D_{s1}(2460)^-$, $e^+e^-\to
D_s^{*+}D_{s1}^*(2536)^-$, $e^+e^-\to D_s^+\bar{D}^0K^-$, $e^+e^-\to
D_s^+\bar{D}^{*0}K^-$, $e^+e^-\to D_s^{*+}\bar{D}^0K^-$, and
$e^+e^-\to D_s^{*+}\bar{D}^{*0}K^-$ processes, where the normalization
factors of the former three processes are estimated according to their
cross sections reported in Refs.~\cite{DsstDsst,DsstDs12460,DsstDsj},
and those of the latter three processes are estimated according to
their cross sections measured in this work. For the shapes of
$M(K^-K^+\pi^+)$ in $B_{{\rm Other\ B.K.G.}}$, we convolve the MC
shape with a Gaussian function. The parameters of the Gaussian
function are fixed to those used in $S_{M(K^-K^+\pi^+)}$.

\begin{figure*}[!htbp]
    \centering
    \includegraphics[width=0.45\textwidth]{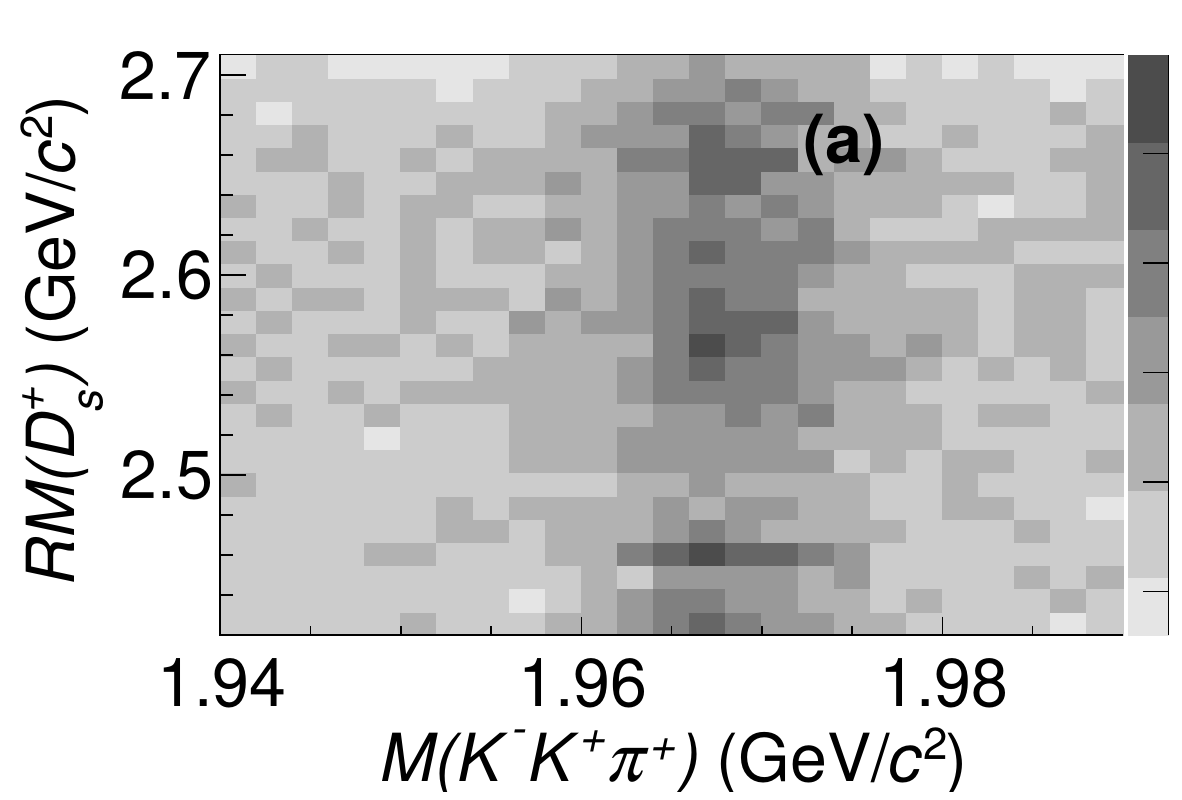}
    \includegraphics[width=0.45\textwidth]{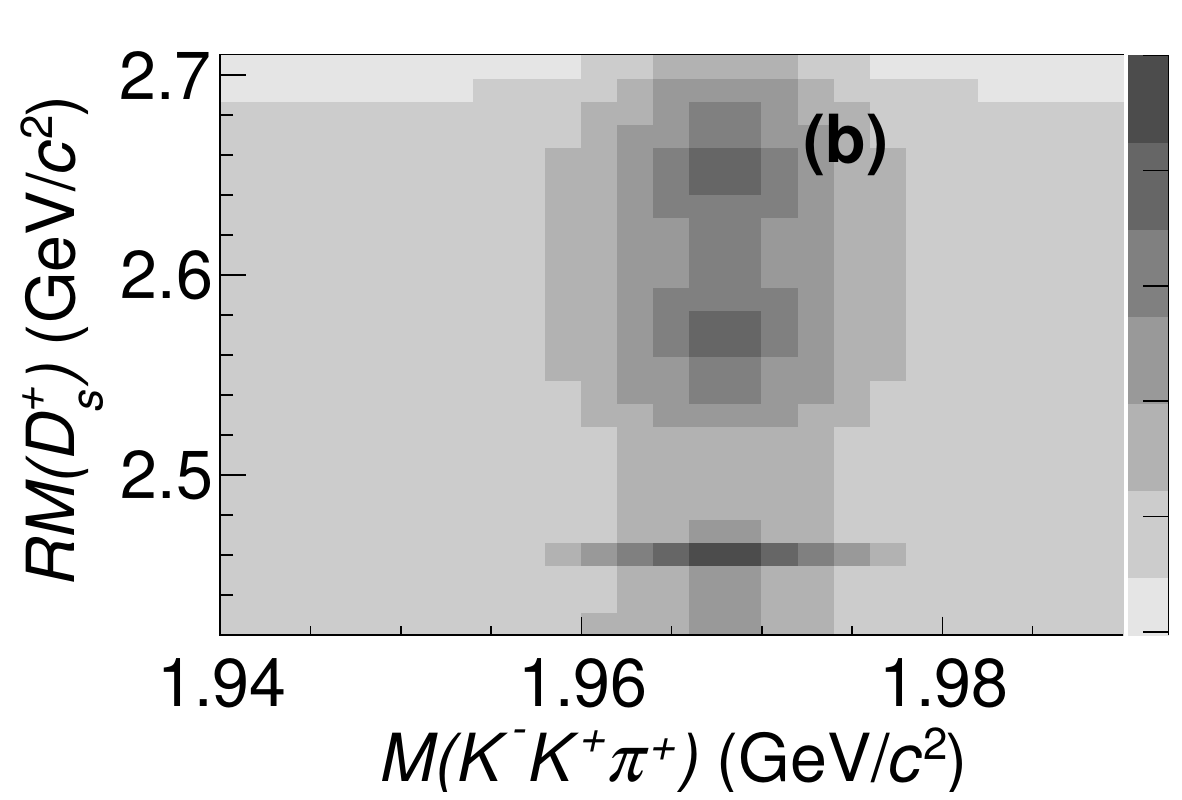}
    \includegraphics[width=0.45\textwidth]{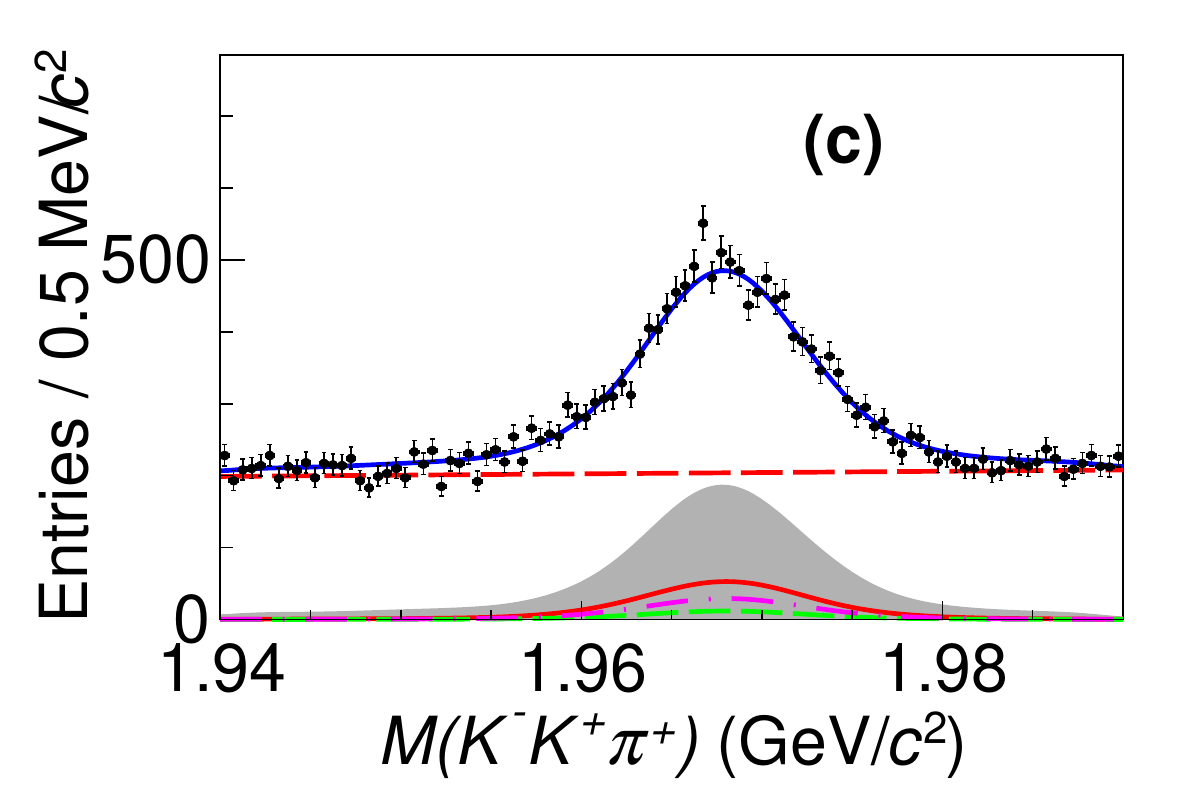}
    \includegraphics[width=0.45\textwidth]{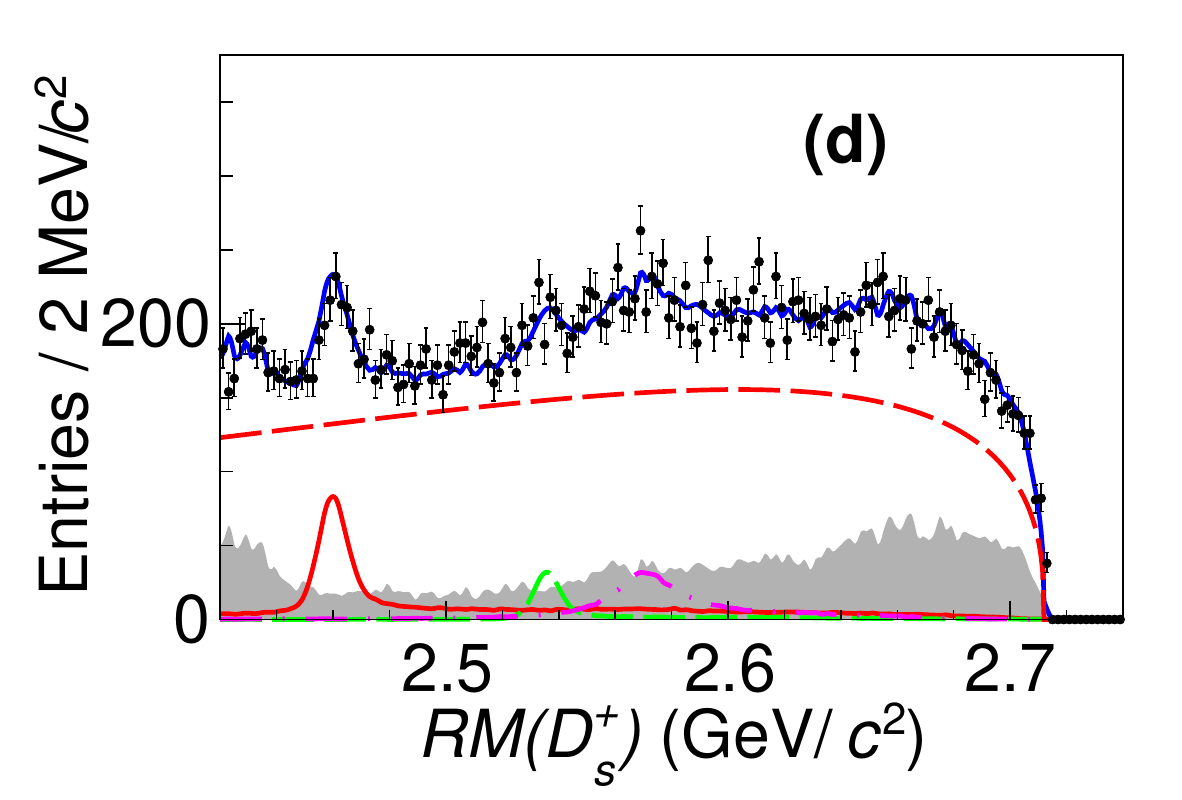}
    \includegraphics[width=0.45\textwidth]{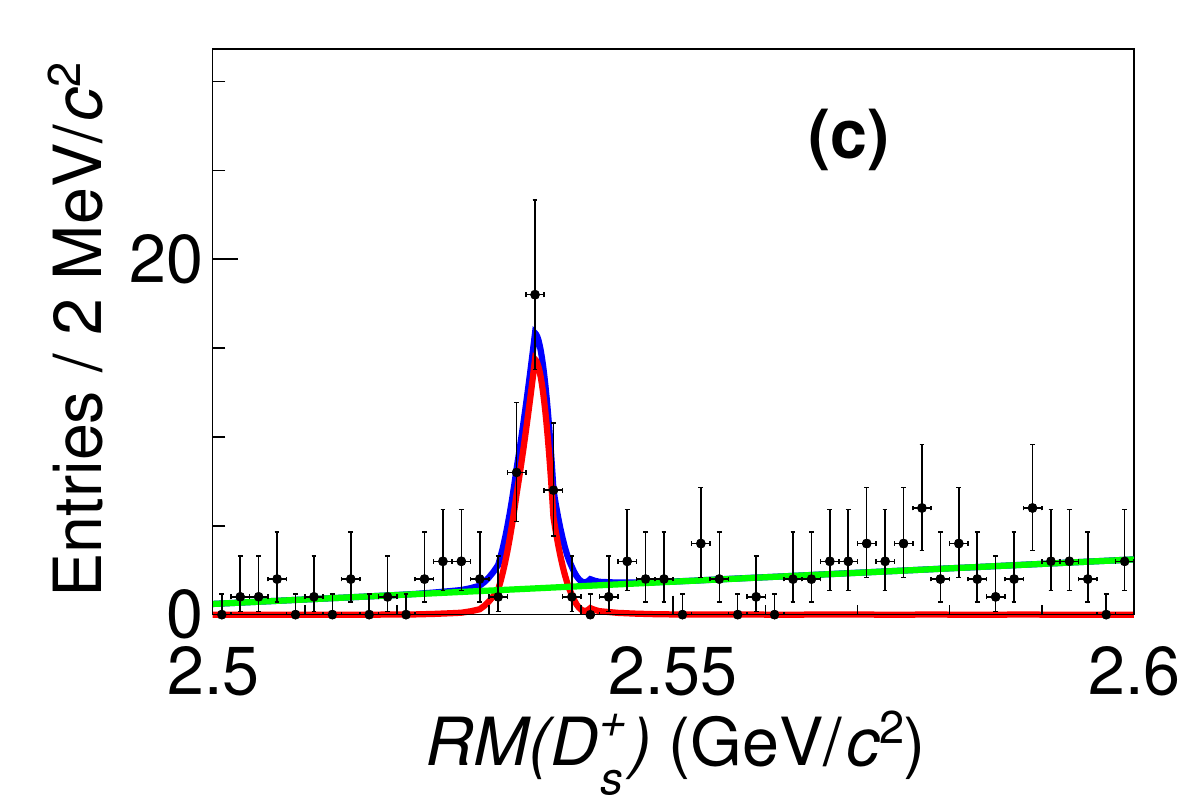}
    \includegraphics[width=0.45\textwidth]{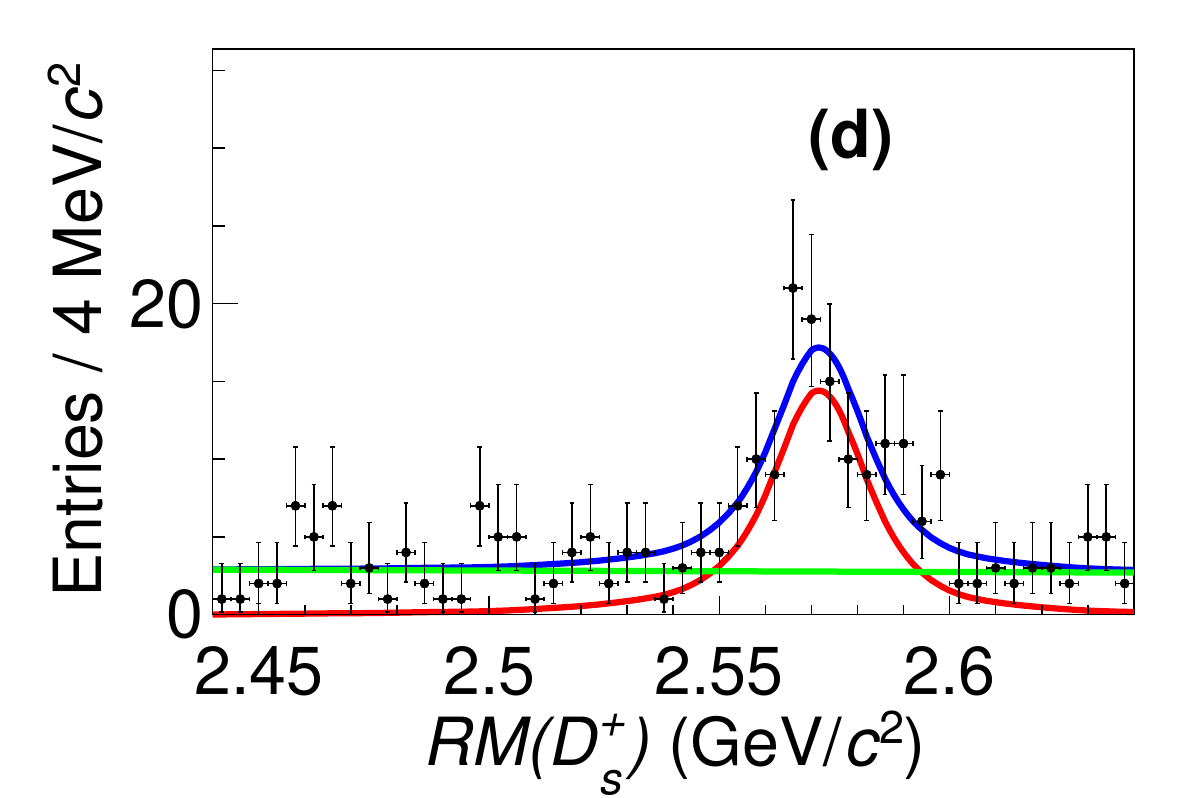}
    \caption{ Top row: Distributions of (a) $RM(D_s^+)$ versus
      $M(K^-K^+\pi^+)$ in data and (b) total fit results of the 2D fit
      in the inclusive analysis. Middle row: Projections of the 2D fit
      in (c) $M(K^-K^+\pi^+)$ and (d) $RM(D_s^+)$ in the inclusive
      analysis. Here, dots with error bars are data, the gray
      histograms are $B_{{\rm Other\ B.K.G.}}$, the red dotted lines
      are $B_{M(K^-K^+\pi^+)}\times B_{RM(D_s^+)}$, the blue solid
      lines are the total fit, and the red solid, green dashed, purple
      dash-dotted lines are $S_{M(K^-K^+\pi^+)}\times S_{RM(D_s^+)}$,
      where $S_{RM(D_s^+)}$ corresponds to the $D_{s1}(2460)^-$,
      $D_{s1}(2536)^-$, and $D^*_{s2}(2573)^-$ MC signal shapes,
      respectively.  Bottom row: Fit results of (e) $RM(D_s^+)$ for
      $D_{s1}(2536)^-$ and (f) $D^*_{s2}(2573)^-$ signals in the
      exclusive analysis. Here, dots with error bars are data, and the
      blue, red, and green solid lines are the total fit, signal shape,
      and background shape, respectively.  }
    \label{pic:inc_fit}
\end{figure*}

In the exclusive analysis, for data samples with $\sqrt{s}\geq4.6$
GeV, the $D_{s1}(2536)^-$ and $D^*_{s2}(2573)^-$ yields are determined
by an extended unbinned likelihood fit of $RM(D_s^+)$.  The signal
shape is described by the MC shape convolved with a Gaussian function,
and the background shape is described by a first-order Chebychev
polynomial function. For the $D_{s1}(2536)^{-}$ signal, the parameters
of the Gaussian function are fixed according to the $RM(D_s^+)$ fit of
all data samples, and for the $D^{*}_{s2}(2573)^{-}$ signal, the
parameters of the Gaussian function are free to vary.  The fit results in
both inclusive and exclusive analyses at $\sqrt{s}=4.680$~GeV are
shown in Fig.~\ref{pic:inc_fit}.  The cross sections of \edos\ and
\edts\ used in the branching fraction measurements are shown in
Tables~\ref{table3} and~\ref{table4}, respectively.  The exclusive
cross sections for \edos\ and \edts\ are shown in
Tables~\ref{table:ex_xs_DsDs1_count} and \ref{table:ex_xs_DsDs1} and
Tables~\ref{table:ex_xs_DsDs2_count} and \ref{table:ex_xs_DsDs2},
respectively.

\begin{table}[htp]
	\centering
	\caption{The measured cross sections for \edos. $\sqrt{s}$, $\epsilon_{{\rm inc}}$, $N_{{\rm inc}}$, $\epsilon_{{\rm exc}}(D_{s}^{+}\rightarrow\skkp)$, $\epsilon_{{\rm exc}}(D_{s}^{+}\rightarrow\sksk)$, $N^{{\rm obs}}_{{\rm exc}}$, $\FS{1}{|1-\Pi|^2}(1+\delta)$, $\sigma^{{\rm inc}}$, and $\sigma^{{\rm exc}}$ represent center-of-mass energy, efficiency in the inclusive analysis, number of events observed in the inclusive analysis, efficiency of $D_{s}^{+}\rightarrow\skkp$ decay mode in the exclusive analysis, efficiency of $D_{s}^{+}\rightarrow\sksk$ decay mode in the exclusive analysis, number of events observed in the exclusive analysis, product of vacuum polarization factor and ISR correction factor, inclusive cross section, and exclusive cross section, respectively. The uncertainties are statistical only.}
	\label{table3}
	\resizebox{\textwidth}{!}{
		\begin{tabular}{ccccccccc}
			\hline\hline
			$\sqrt{s}$ (GeV) & $\epsilon_{{\rm inc}}$ (\%) & $N_{{\rm inc}}$ & $\epsilon_{{\rm exc}}(D_{s}^{+}\rightarrow\skkp)$ (\%) & $\epsilon_{{\rm exc}}(D_{s}^{+}\rightarrow\sksk)$ (\%) &  $N^{{\rm obs}}_{{\rm exc}}$ & $\FS{1}{|1-\Pi|^2}(1+\delta)$ & $\sigma^{{\rm inc}}$(pb) & $\sigma^{{\rm exc}}$(pb)\\
			\hline
			4.600 & 30.2 & 159.5$\pm$25.2 & 9.3  & 13.6 & 28.9$\pm$5.7  & 0.79 & 21.1$\pm$3.3  & 9.7 $\pm$1.9 \\
			4.620 & 29.1 & 113.1$\pm$26.0 & 7.9  & 11.1 & 16.2$\pm$4.6  & 0.95 & 14.5$\pm$3.3  & 6.1 $\pm$1.7 \\
			4.640 & 29.3 & 139.6$\pm$27.0 & 8.1  & 11.3 & 17.3$\pm$4.4  & 0.88 & 18.3$\pm$3.5  & 6.5 $\pm$1.6 \\
			4.660 & 29.2 & 111.4$\pm$27.7 & 7.6  & 10.4 & 11.6$\pm$4.0  & 1.02 & 13.1$\pm$3.3  & 4.2 $\pm$1.4 \\
			4.680 & 29.3 & 289.5$\pm$50.0 & 7.4  & 10.2 & 30.4$\pm$6.1  & 1.02 & 10.7$\pm$1.9  & 3.5 $\pm$0.7 \\
			4.700 & 29.3 & 126.3$\pm$28.6 & 7.8  & 10.6 & 10.4$\pm$3.7  & 1.0  & 15.0$\pm$3.4  & 3.7 $\pm$1.3 \\
			\hline\hline
		\end{tabular}
	}
\end{table}

\begin{table}[htp]
	\centering
	\caption{The measured cross sections for \edts. Definitions of each symbol are the same as those in Table \ref{table3}.}
	\label{table4}
	\resizebox{\textwidth}{!}{
		\begin{tabular}{ccccccccc}
			\hline\hline
			$\sqrt{s}$ (GeV) & $\epsilon_{{\rm inc}}$ (\%) & $N_{{\rm inc}}$ & $\epsilon_{{\rm exc}}(D_{s}^{+}\rightarrow\skkp)$ (\%) & $\epsilon_{{\rm exc}}(D_{s}^{+}\rightarrow\sksk)$ (\%) &  $N^{{\rm obs}}_{{\rm exc}}$ & $\FS{1}{|1-\Pi|^2}(1+\delta)$ & $\sigma^{{\rm inc}}$(pb) & $\sigma^{{\rm exc}}$(pb)\\
			\hline
			4.600 & 30.2 & 341.8$\pm$43.7 & 15.6 & 13.6 & 94.8$\pm$11.0  & 0.76 & 47.0$\pm$6.0  & 19.7$\pm$2.3 \\
			4.620 & 29.2 & 381.8$\pm$44.3 & 14.3 & 11.1 & 96.1$\pm$11.1  & 0.77 & 60.5$\pm$7.0  & 24.5$\pm$2.8 \\
			4.640 & 29.4 & 356.7$\pm$46.7 & 13.1 & 11.3 & 71.5$\pm$10.1  & 0.95 & 42.9$\pm$5.6  & 15.1$\pm$2.1 \\
			4.660 & 29.2 & 266.5$\pm$44.6 & 12.6 & 10.4 & 59.3$\pm$10.5  & 0.88 & 36.2$\pm$6.1  & 14.6$\pm$2.6 \\
			4.680 & 29.4 & 721.0$\pm$86.7 & 11.1 & 10.2 & 109.6$\pm$14.6  & 1.06 & 25.7$\pm$3.1  & 8.1 $\pm$1.1 \\
			4.700 & 29.3 & 158.8$\pm$53.4 & 9.3  & 10.6 & 19.1$\pm$7.0  & 1.32 & 14.2$\pm$4.8  & 4.2 $\pm$1.5 \\
			\hline\hline
		\end{tabular}
	}
\end{table}

\section{Absolute branching fraction measurement}
Formulas of the likelihood $L^{{\rm inc}, {\rm exc}}_{i,j}$ is given as
\begin{equation}
    L^{{\rm inc}}_{i,j}(\sigma_{i,j}^{{\rm inc}},\delta_{i,j}^{{\rm inc}},\sigma_{i,j})=\FS{1}{\sqrt{2\pi}\delta_{i,j}^{{\rm inc}}}e^{\rule{0.215cm}{0.4pt}\FS{(\sigma_{i,j}^{{\rm inc}}-\sigma_{i,j})^2}{2(\delta_{i,j}^{{\rm inc}})^2}}\label{2},
\end{equation}
\begin{equation}
    L_{i,j}^{{\rm exc}}(\sigma_{i,j}^{{\rm exc}},\delta_{i,j}^{{\rm exc}},\sigma_{i,j},\mathcal{B}_{i})=\FS{1}{\sqrt{2\pi}\delta_{i,j}^{{\rm exc}}}e^{\rule{0.215cm}{0.4pt}\FS{(\sigma_{i,j}^{{\rm exc}}-\sigma_{i,j} \mathcal{B}_{i})^2}{2(\delta_{i,j}^{{\rm exc}})^2}}\label{3}.
\end{equation}

\section{Fit to measured cross section}
To study the resonance structures in the \edos\ and \edts\ processes, least-$\chi^2$ fits to the measured cross sections are performed. The cross sections are described with the coherent sum of two constant width Breit-Wigner ($BW$) functions,
\begin{equation}
    \sigma(\sqrt{s})=|BW_0(\sqrt{s})+BW_1(\sqrt{s})e^{i\phi_1}|^2.
\end{equation}

\begin{equation}
	BW_k(\sqrt{s})=\frac{\sqrt{12\pi\Gamma_k^{{\rm tot}}\Gamma_k^{e^+e^-}\mathcal{B}_k}}{s-M_k^2+iM_k\Gamma_k^{{\rm tot}}}\frac{\sqrt{\Phi(\sqrt{s})}}{\sqrt{\Phi(M_k)}},
\end{equation}
\begin{equation}
    \Phi(\sqrt{s})=\frac{q(\sqrt{s})^{2l+1}}{s},
\end{equation}
where $M_k$, $\Gamma_k^{{\rm tot}}$, and $\Gamma_k^{e^+e^-}$ are the mass, width, and electronic partial width of the $k^{\rm th}$ structure ($R_k$), respectively; $\mathcal{B}_{k}$ is the branching fraction of the decay $R_k\to  D_s^+D_{s1}(2536)^{-}/D_s^+D^{*}_{s2}(2573)^{-}$, $\phi$ is the relative phase between $R_1$ and $R_0$, $q(\sqrt{s})$ is the momentum of $D_s^+$ in the rest frame of $R_k$, $\Phi(\sqrt{s})$ is the phase space factor of the two-body decay, and $l$ is the angular momentum of $R_k\to  D_s^+D_{s1}(2536)^{-}/D_s^+D^{*}_{s2}(2573)^{-}$. 

The angular distribution of the \edos\ process is described by $1+\alpha {\rm cos}^2\theta$, where $\alpha=-0.65\pm0.22$ is measured with our data. This indicates both $S$-wave and $D$-wave contributions exist, however, their fractions cannot be determined precisely due to the low statistics. We assume a pure $D$-wave process in the nominal fit and take a pure $S$-wave alternatively to estimate the systematic uncertainty. The \edts\ is pure $D$-wave.

In the analysis of \edos, the mass and width of $BW_0$ are determined
to be $M_0=(4584\pm14\pm68)$~MeV/$c^2$ and $\Gamma^{{\rm tot}}_0=(57\pm
12\pm211)$~MeV and those of $BW_1$ are
$M_1=(4749.9\pm8.2\pm6.7)$~MeV/$c^2$ and
$\Gamma^{{\rm tot}}_1=(24.9\pm8.0\pm7.8)$~MeV, where the first uncertainties are
statistical and the second systematic.  In the analysis of \edts, the
mass and width of $BW_0$ are determined to be $M_0=(4603.1\pm
3.9\pm0.8)$~MeV/$c^2$ and $\Gamma^{{\rm tot}}_0=(45.2\pm5.7\pm0.7)$~MeV and those
of $BW_1$ are $M_1=(4720\pm13\pm2)$~MeV/$c^2$ and
$\Gamma^{{\rm tot}}_1=(50\pm12\pm1)$~MeV.

\section{Systematic uncertainty}

\subsection{Absolute branching fraction measurement}
The systematic uncertainties from $B_{{\rm Other\ B.K.G.}}$ in the 2D
fit are estimated as 2.1\% and 3.0\% for $\dod$ and $\dtd$,
respectively, by sampling the cross sections of the process in
$B_{{\rm Other\ B.K.G.}}$ 500 times according to their uncertainties.

The systematic uncertainties from fit range of $RM(D_s^+)$ in
inclusive (exclusive) analyses are estimated by the
``Barlow-test''~\cite{barlow1,barlow2} as 0.1\% and 0.7\% (1.1\% and
0.8\%) for $\dod$ and $\dtd$, respectively.

The parameters of the convolved Gaussian function in the 2D fit are
changed by one standard deviation, and the largest changes are taken as
the systematic uncertainty as 2.5\% and 3.6\% for $\dod$ and $\dtd$,
respectively.

The background shape of $M(K^-K^+\pi^+)$ in the 2D fit is changed from
the first order Chebychev polynomial function to a second order one,
and the systematic uncertainties are estimated as 5.0\% and 7.4\% for
$\dod$ and $\dtd$, respectively.

The background shape of $RM(D_s^+)$ in the 2D fit is changed from an
ARGUS function to $f(M) = (M - M_a)^c(M_b - M)^d$, where $c$ and $d$
are floating parameters, and $M_a$ ($M_b$) is the lower (upper) limit of
mass distribution, which is fixed as $M_a=0$
($M_{b}=\sqrt{s}-m_{D_s^+}$)~\cite{DsstDs12460}.  The systematic
uncertainties are estimated as 1.0\% and 1.6\% for $\dod$ and $\dtd$,
respectively.

\subsection{Exclusive cross section measurement}
The systematic uncertainties of the $RM(D_s^+)$ fit range are estimated
the same way as in the estimation of the absolute branching fraction
measurements as 1.7\% and 0.3\% for \edos\ and \edts, respectively.

The background shape of $RM(D_s^+)$ is changed from the first order
Chebychev polynomial function to a second order one.

All the systematic uncertainties are shown in Tables~\ref{table:sys_ex_xs_DsDs1_1} and~\ref{table:sys_ex_xs_DsDs2_1}.

\subsection{Fit to measured cross section}

The systematic uncertainties in the resonance parameters of $R_0$ and
$R_1$ mainly stem from the center-of-mass energy measurement,
center-of-mass energy spread, and systematic uncertainty of the cross
sections.

The center-of-mass energies of data samples with $\sqrt{s}<4.61$~GeV are measured with dimuon events with an uncertainty of $\pm0.8$~MeV~\cite{XYZEcm1}, while those with $\sqrt{s}>4.6$~GeV are measured with $\Lambda_c^+\bar{\Lambda}_c^-$ events with an uncertainty of $\pm0.6$~MeV~\cite{XYZEcm2}. Thus 0.8~MeV is taken as the systematic uncertainty, and propagates to the masses of $BW_0$ and $BW_1$ by the same amount.

The systematic uncertainty from the center-of-mass energy spread is
estimated by convolving the fit formula with a Gaussian function with
a width of 1.6~MeV, which is the energy spread determined from the
measurement results of the Beam Energy Measurement System~\cite{bems}.

The systematic uncertainties from the cross section measurement uncommon among data samples will influence the masses and widths of $R_0$ and $R_1$, which include ISR correction, background shape of $RM(D_s^+)$, and tracking and PID efficiencies of $K^-$ not from $D_s^+$.  The corresponding systematic uncertainty is estimated by including the uncertainty in the fit to the cross section, and taking the differences on the parameters as the systematic uncertainties. 

The fraction of the $S$-wave and $D$-wave in the \edos\ decay also has influence on the results; conservatively, we assume $S$-wave is dominant, and the difference from the nominal fit is taken as the systematic uncertainty.

\begin{table}[htp]
	\centering
	\caption{Relative systematic uncertainty (\%) 
	in the exclusive cross section measurement for \edos.}
	\label{table:sys_ex_xs_DsDs1_1}
		\begin{tabular}{lccccccccccccc}
			\hline\hline
			$\sqrt{s}$ (GeV) & 4.600 & 4.610 & 4.620 & 4.640 & 4.660 & 4.680 & 4.700 & 4.740 & 4.750 & 4.780 & 4.840 & 4.914 & 4.946 \\
			\hline
			Fit range (exc)                   & 1.7  & 1.7  & 1.7  & 1.7  & 1.7  & 1.7  & 1.7  & 1.7  & 1.7  & 1.7  & 1.7  & 1.7  & 1.7  \\
			Mass window ($M(D_s^+)$)         & 3.4  & 3.4  & 3.4  & 3.4  & 3.4  & 3.4  & 3.4  & 3.4  & 3.4  & 3.4  & 3.4  & 3.4  & 3.4  \\
			Mass window ($RM(D_s^+K^-)$)     & 4.3  & 4.3  & 4.3  & 4.3  & 4.3  & 4.3  & 4.3  & 4.3  & 4.3  & 4.3  & 4.3  & 4.3  & 4.3  \\
			Tracking \& PID ((tracks from $D_s^+$))                  & 1.9  & 1.9  & 1.9  & 1.9  & 1.9  & 1.9  & 1.9  & 1.9  & 1.9  & 1.9  & 1.9  & 1.9  & 1.9  \\
			Tracking \& PID ($K^{\pm}$ not from $D_s^+$) & 1.7  & 1.5  & 1.4  & 1.6  & 1.3  & 1.2  & 1.1  & 0.9  & 0.9  & 0.8  & 0.8  & 0.8  & 0.8  \\
			$\mathcal{B}(\dskkp)$                     & 1.9  & 1.9  & 1.9  & 1.9  & 1.9  & 1.9  & 1.9  & 1.9  & 1.9  & 1.9  & 1.9  & 1.9  & 1.9 \\
			$\mathcal{B}(\ksk)$                       & 2.4  & 2.4  & 2.4  & 2.4  & 2.4  & 2.4  & 2.4  & 2.4  & 2.4  & 2.4  & 2.4  & 2.4  & 2.4  \\
			B.K.G. shape (exc: $RM(D_s^+)$)   & 4.7  & 3.6  & 9.2  & 6.1  & 5.9  & 0.4  & 0.6  & 0.4  & 8.1  & 6.0  & 6.8  & 8.2  & 13.4 \\
			VVS\_PWave                       & 0.2  & 0.2  & 0.2  & 0.2  & 0.2  & 0.2  & 0.2  & 0.2  & 0.2  & 0.2  & 0.2  & 0.2  & 0.2  \\
			{\sc AngSam}                           & 3.3  & 3.3  & 3.3  & 3.3  & 3.3  & 3.3  & 3.3  & 3.3  & 3.3  & 3.3  & 3.3  & 3.3  & 3.3  \\
			ISR                              & 1.4  & 3.5  & 3.6  & 1.9  & 2.2  & 1.7  & 2.0  & 3.2  & 1.1  & 1.0  & 0.9  & 0.4  & 2.3  \\
			Luminosity                       & 1.0  & 1.0  & 1.0  & 1.0  & 1.0  & 1.0  & 1.0  & 1.0  & 1.0  & 1.0  & 1.0  & 1.0  & 1.0  \\
			Total                            & 9.2  & 9.2  & 12.5 & 10.1 & 10.0 & 7.9  & 8.0  & 8.3  & 11.2 & 9.8  & 10.3 & 11.2 & 15.6 \\
			\hline\hline
		\end{tabular}
\end{table}

\begin{table}[htp]
	\centering
	\caption{Relative systematic uncertainty (\%) 
	in the exclusive cross section measurement for \edts.}
	\label{table:sys_ex_xs_DsDs2_1}
		\begin{tabular}{lccccccccccccc}
			\hline\hline
			$\sqrt{s}$ (GeV) & 4.600 & 4.610 & 4.620 & 4.640 & 4.660 & 4.680 & 4.700 & 4.740 & 4.750 & 4.780 & 4.840 & 4.914 & 4.946 \\
			\hline
			Fit range (exc)                   & 0.3  & 0.3  & 0.3  & 0.3  & 0.3  & 0.3  & 0.3  & 0.3  & 0.3  & 0.3  & 0.3  & 0.3  & 0.3  \\
			Mass window ($M(D_s^+)$)         & 5.5  & 5.5  & 5.5  & 5.5  & 5.5  & 5.5  & 5.5  & 5.5  & 5.5  & 5.5  & 5.5  & 5.5  & 5.5  \\
			Mass window ($RM(D_s^+K^-)$)     & 4.3  & 4.3  & 4.3  & 4.3  & 4.3  & 4.3  & 4.3  & 4.3  & 4.3  & 4.3  & 4.3  & 4.3  & 4.3  \\
			Tracking \& PID ((tracks from $D_s^+$))              & 1.9  & 1.9  & 1.9  & 1.9  & 1.9  & 1.9  & 1.9  & 1.9  & 1.9  & 1.9  & 1.9  & 1.9  & 1.9  \\
			Tracking \& PID ($K^{\pm}$ not from $D_s^+$) & 0.1  & 0.1  & 0.0  & 0.0  & 0.0  & 0.1  & 0.1  & 0.1  & 0.0  & 0.1  & 0.0  & 0.1  & 0.1  \\
			$\mathcal{B}(\dskkp)$                     & 1.9  & 1.9  & 1.9  & 1.9  & 1.9  & 1.9  & 1.9  & 1.9  & 1.9  & 1.9  & 1.9  & 1.9  & 1.9  \\
			$\mathcal{B}(\ksk)$                       & 2.4  & 2.4  & 2.4  & 2.4  & 2.4  & 2.4  & 2.4  & 2.4  & 2.4  & 2.4  & 2.4  & 2.4  & 2.4  \\
			B.K.G. shape (exc: $RM(D_s^+)$)   & 3.0  & 6.5  & 5.4  & 13.4 & 5.5  & 8.8  & 7.0  & 2.5  & 4.1  & 7.2  & 13.0 & 46.3 & 25.8 \\
			ISR                              & 1.0  & 2.7  & 1.1  & 1.3  & 0.8  & 1.2  & 2.9  & 5.3  & 2.5  & 0.7  & 0.6  & 0.6  & 1.6  \\
			Luminosity                       & 1.0  & 1.0  & 1.0  & 1.0  & 1.0  & 1.0  & 1.0  & 1.0  & 1.0  & 1.0  & 1.0  & 1.0  & 1.0  \\
			Total                            & 8.5  & 10.6 & 9.7  & 15.6 & 9.7  & 11.9 & 11.0 & 9.9  & 9.3  & 10.7 & 15.2 & 47.0 & 27.0 \\
			\hline\hline
		\end{tabular}
\end{table}

\begin{table}[htp]
	\centering
	\caption{The measured cross sections for \edos\ in the exclusive analysis by the counting method. $N^{{\rm count}}_{{\rm exc}}$ is the number of events obtained by the counting method. The uncertainties for $N^{{\rm count}}_{{\rm exc}}$ and $\sigma^{{\rm exc}}$ are statistical only.}
	\label{table:ex_xs_DsDs1_count}
		\begin{tabular}{cccccc}
			\hline\hline
			$\sqrt{s}$ (GeV) & $\epsilon_{{\rm exc}}(D_{s}^{+}\rightarrow\skkp)$ (\%) & $\epsilon_{{\rm exc}}(D_{s}^{+}\rightarrow\sksk)$ (\%) & $N^{{\rm count}}_{{\rm exc}}$ & $\FS{1}{|1-\Pi|^2}(1+\delta)$ & $\sigma^{{\rm exc}}$(pb)\\
			\hline
			4.530 & 10.2 & 15.1 & $3.0_{-14.5}^{+46.5}$ & 0.7  & $0.6_{-2.8}^{+9.0} $\\
			4.575 & 9.4  & 13.8 & $0.3_{-0.0}^{+27.5}$ & 0.8  & $0.1_{-0.0}^{+11.2}$\\
			\hline\hline
		\end{tabular}
\end{table}

\begin{table}[htp]
	\centering
	\caption{The measured cross sections for \edos\ in the exclusive analysis. The first uncertainties for $N^{{\rm obs}}_{{\rm exc}}$ and $\sigma^{{\rm exc}}$ are statistical and the second ones for $\sigma^{{\rm exc}}$ are systematic.}
	\label{table:ex_xs_DsDs1}
		\begin{tabular}{cccccc}
			\hline\hline
			$\sqrt{s}$ (GeV) & $\epsilon_{{\rm exc}}(D_{s}^{+}\rightarrow\skkp)$ (\%) & $\epsilon_{{\rm exc}}(D_{s}^{+}\rightarrow\sksk)$ (\%) & $N^{{\rm obs}}_{{\rm exc}}$ & $\FS{1}{|1-\Pi|^2}(1+\delta)$ & $\sigma^{{\rm exc}}$(pb)\\
			\hline
			4.600 & 9.4  & 13.6 & 28.9$\pm$5.7  & 0.79 & 9.7 $\pm$1.9 $\pm$0.9 \\
			4.610 & 9.3  & 12.2 & 4.6 $\pm$2.3  & 0.83 & 9.1 $\pm$4.4 $\pm$0.8 \\
			4.620 & 8.7  & 11.1 & 16.2$\pm$4.6  & 0.95 & 6.1 $\pm$1.7 $\pm$0.8 \\
			4.640 & 7.9  & 11.3 & 17.3$\pm$4.4  & 0.88 & 6.5 $\pm$1.6 $\pm$0.7 \\
			4.660 & 8.1  & 10.4 & 11.6$\pm$4.0  & 1.02 & 4.2 $\pm$1.4 $\pm$0.4 \\
			4.680 & 7.6  & 10.2 & 30.4$\pm$6.1  & 1.02 & 3.5 $\pm$0.7 $\pm$0.3 \\
			4.700 & 7.4  & 10.6 & 10.4$\pm$3.7  & 1.0  & 3.7 $\pm$1.3 $\pm$0.3 \\
			4.740 & 7.8  & 12.8 & 7.5 $\pm$3.0  & 0.95 & 7.5 $\pm$3.0 $\pm$0.6 \\
			4.750 & 9.5  & 14.0 & 31.6$\pm$5.9  & 0.71 & 17.6$\pm$3.3 $\pm$2.0 \\
			4.780 & 10.2 & 12.0 & 26.7$\pm$5.5  & 0.96 & 9.2 $\pm$1.9 $\pm$0.9 \\
			4.840 & 8.8  & 10.3 & 13.9$\pm$4.0  & 1.12 & 4.5 $\pm$1.3 $\pm$0.5 \\
			4.914 & 7.9  & 12.2 & 11.1$\pm$3.6  & 0.88 & 9.5 $\pm$3.1 $\pm$1.1 \\
			4.946 & 9.6  & 9.8  & 4.4 $\pm$2.4  & 1.15 & 4.6 $\pm$2.5 $\pm$0.7 \\
			\hline\hline
		\end{tabular}
\end{table}

\begin{table}[htp]
	\centering
	\caption{The measured cross sections for \edts\ in the exclusive analysis by the counting method. $N^{{\rm count}}_{{\rm exc}}$ is the number of events obtained by the counting method.  The uncertainties for $N^{{\rm count}}_{{\rm exc}}$ and $\sigma^{{\rm exc}}$ are statistical only.}
	\label{table:ex_xs_DsDs2_count}
		\begin{tabular}{cccccc}
			\hline\hline
			$\sqrt{s}$ (GeV) & $\epsilon_{{\rm exc}}(D_{s}^{+}\rightarrow\skkp)$ (\%) & $\epsilon_{{\rm exc}}(D_{s}^{+}\rightarrow\sksk)$ (\%) & $N^{{\rm count}}_{{\rm exc}}$ & $\FS{1}{|1-\Pi|^2}(1+\delta)$ & $\sigma^{{\rm exc}}$(pb)\\
			\hline
			4.575 & 16.2 & 24.5 & $-0.5_{-0.0}^{+10.1}$ & 0.7  & $-0.2_{-0.0}^{+4.3} $\\
			\hline\hline
		\end{tabular}
\end{table}

\begin{table}[htp]
	\centering
	\caption{The measured cross sections for \edts\ process in the exclusive analysis. The first uncertainties for $N^{{\rm obs}}_{{\rm exc}}$ and $\sigma^{{\rm exc}}$ are statistical and the second ones for $\sigma^{{\rm exc}}$ are systematic.}
	\label{table:ex_xs_DsDs2}
		\begin{tabular}{cccccc}
			\hline\hline
			$\sqrt{s}$ (GeV) & $\epsilon_{{\rm exc}}(D_{s}^{+}\rightarrow\skkp)$ (\%) & $\epsilon_{{\rm exc}}(D_{s}^{+}\rightarrow\sksk)$ (\%) & $N^{{\rm obs}}_{{\rm exc}}$ & $\FS{1}{|1-\Pi|^2}(1+\delta)$ & $\sigma^{{\rm exc}}$(pb)\\
			\hline
			4.600 & 15.6 & 13.6 & 94.8$\pm$11.0 & 0.79 & 19.7$\pm$2.3 $\pm$1.7 \\
			4.610 & 14.4 & 12.2 & 15.9$\pm$5.4  & 0.83 & 18.5$\pm$6.3 $\pm$2.0 \\
			4.620 & 14.3 & 11.1 & 96.1$\pm$11.1 & 0.95 & 24.5$\pm$2.8 $\pm$2.4 \\
			4.640 & 13.1 & 11.3 & 71.5$\pm$10.1 & 0.88 & 15.1$\pm$2.1 $\pm$2.4 \\
			4.660 & 12.6 & 10.4 & 59.3$\pm$10.5 & 1.02 & 14.6$\pm$2.6 $\pm$1.4 \\
			4.680 & 11.1 & 10.2 & 109.6$\pm$14.6 & 1.02 & 8.1 $\pm$1.1 $\pm$1.0 \\
			4.700 & 9.3  & 10.6 & 19.1$\pm$7.0  & 1.0  & 4.2 $\pm$1.5 $\pm$0.5 \\
			4.740 & 9.7  & 12.8 & 4.7 $\pm$3.1  & 0.95 & 3.3 $\pm$2.1 $\pm$0.3 \\
			4.750 & 11.8 & 14.0 & 18.5$\pm$6.0  & 0.71 & 7.1 $\pm$2.3 $\pm$0.7 \\
			4.780 & 12.3 & 12.0 & 36.4$\pm$8.0  & 0.96 & 9.5 $\pm$2.1 $\pm$1.0 \\
			4.840 & 11.1 & 10.3 & 29.3$\pm$6.9  & 1.12 & 7.4 $\pm$1.7 $\pm$1.1 \\
			4.914 & 11.6 & 12.2 & 15.9$\pm$5.5  & 0.88 & 10.7$\pm$3.7 $\pm$5.0 \\
			4.946 & 10.4 & 9.8  & 9.0 $\pm$4.5  & 1.15 & 7.7 $\pm$3.9 $\pm$2.1 \\
			\hline\hline
		\end{tabular}
\end{table}

The systematic uncertainties in the measured resonance parameters are shown in Tables~\ref{table:sys_par_DsDs1} $\sim$ \ref{table:sys_par_DsDs2}.

\begin{table}[htp]
	\centering
	\caption{The systematic uncertainties in the measurement of the resonance parameters in the \edos\ process. $\sqrt{s}$ represents the systematic uncertainty from the center-of-mass energy measurement. Cross Section represents the systematic uncertainty from the cross section measurements which are uncommon among data samples. $S$-wave represents the systematic uncertainty from the $S$-wave decay of $D_s^+D_{s1}(2536)^{-}$. The units of $M_{0}$ and $\Gamma_{0}^{{\rm tot}}$ are MeV/$c^{2}$ and MeV, respectively.}
	\begin{tabular}{lcccccc}
		\hline\hline
		& $\sqrt{s}$ & Center-of-mass Energy Spread & Cross Section & $S$-wave & Overall  &\\
		\hline
		$M_{0}$            & 0.8        & 7.3             & 7.8          & 67.1         & 67.9    &\\
		$\Gamma_{0}^{{\rm tot}}$ & -        & 3.9             & 3.7          & 211.3         & 211.4    &\\
		$M_{1}$            & 0.8        & 4.7             & 4.7          & 0.5          & 6.7     &\\
		$\Gamma_{1}^{{\rm tot}}$ & -        & 3.5             & 4.1          & 5.7          & 7.8     &\\
		\hline\hline
	\end{tabular}
	\label{table:sys_par_DsDs1}
\end{table}

\begin{table}[htp]
	\centering
	\caption{The systematic uncertainties in the measurement of the resonance parameters in the \edts\ process. $\sqrt{s}$ represents the systematic uncertainty from the center-of-mass energy measurement. Cross Section represents the systematic uncertainty from the cross section measurements which are uncommon among data samples. The units of $M_{1}$ and $\Gamma_{1}^{{\rm tot}}$ are MeV/$c^{2}$ and MeV, respectively.}
	\begin{tabular}{lccccc}
		\hline\hline
		& $\sqrt{s}$ & Center-of-mass Energy Spread & Cross Section & Overall  &\\
		\hline
		$M_{0}$            & 0.8        & 0.0              & 0.1          & 0.8     &\\
		$\Gamma_{0}^{{\rm tot}}$ & -        & 0.2              & 0.7          & 0.7     &\\
		$M_{1}$            & 0.8        & 0.2              & 1.3          & 1.5     &\\
		$\Gamma_{1}^{{\rm tot}}$ & -        & 0.2              & 0.9          & 0.9     &\\
		\hline\hline
	\end{tabular}
	\label{table:sys_par_DsDs2}
\end{table}